\documentclass[preprint, tighten, twocolumn]{aastex63}

\usepackage[clockwise,figuresright]{rotating}

\usepackage{amsfonts,amsmath,amssymb}

\usepackage{lineno}

\newcommand{\W}{\mbox{$\mathcal{W}$}}

\received{}
\revised{}
\accepted{}

\submitjournal{ApJ}

\shorttitle{Winds Driven by LF and MF}
\shortauthors{Yang et al.}

\begin{document}

\title{Magnetohydrodynamic winds driven by line force from the standard thin disk around supermassive black holes: II. a possible model for ultra-fast outflows in radio-loud AGNs}


\author[0000-0002-2419-9590]{Xiao-Hong Yang}

\affiliation{Department of Physics, Chongqing University, Chongqing 400044, People's Republic of China; yangxh@cqu.edu.cn}

\begin{abstract}
In radio-loud active galactic nuclei (AGNs), ultra-fast outflows (UFOs) were detected at the inclination angle of $\sim10^{\rm o}$--$70^{\rm o}$ away from jets. Except for the inclination angle of UFOs, the UFOs in radio-loud AGNs have similar properties to that in radio-quiet AGNs. The UFOs with such low inclination cannot be explained in the line-force mechanism. The magnetic-driving mechanism is suggested to explain the UFOs based on a self-similar solution with radiative transfer calculations. However, the energetics of self-similar solution need to be further confirmed based on numerical simulations. To understand the formation and acceleration of UFOs in radio-loud AGNs, this paper presents a model of the disk winds driven by both line force and magnetic field and implements numerical simulations. Initially, a magnetic field is set to 10 times stronger than the gas pressures at the disk surface. Simulation results imply that the disk winds driven by both line force and magnetic field could describe the properties of UFOs in radio-loud AGNs. Pure magnetohydrodynamics (MHDs) simulation is also implemented. When the initial conditions are the same, the hybrid models of magnetic fields and line force are more helpful to form UFOs than the pure MHD models. It is worth studying the case of a stronger magnetic field to confirm this result.
\end{abstract}


\section{Introduction} \label{sec:intro}
The key features of outflows, such as the blue-shifted lines, are frequently detected on the UV, soft X-ray ($<$3 keV) and hard X-ray (5--10 keV) spectra of active galactic nuclei (AGNs, e.g. Hamann et al. 1997; 2018; Misawa 2007; Mckernan et al. 2007; Tombesi et al. 2010; 2012; Gofford et al. 2013; 2015). Among the blue-shifted lines, the highly ionized Fe XXV and Fe XXVI absorption lines are also observed at the hard X-ray band, based on XMM-Newton and Suzakue observation (Tombesi et al. 2010; 2012; Gofford et al. 2013; 2015). The Fe XXV and Fe XXVI absorption lines imply that a highly ionized absorber fast moves away from its nucleus at $\sim$0.03--0.3 $c$ (Tombesi et al. 2012). Tombesi et al. (2010) first used ultra-fast outflows (UFOs) to define the highly-ionized absorbers that move outwards at velocities higher than 10$^4$ km s$^{-1}$. Tombesi et al. (2011) implemented a photoionization modeling to derive the basic properties of UFOs and found that the ionization parameter ($\xi$) and column density ($N_{\rm H}$) of UFOs are distributed in the range of log($\xi$/(erg s$^{-1}$ cm))$\sim$3--6 and $10^{22}$ cm$^{-2}$$\lesssim N_{\rm H}\lesssim$ $10^{24}$ cm$^{-2}$, respectively (Tombesi et al. 2011). It is estimated that the UFOs are formed in the interval of $10^2$--$10^4$ Schwarzschild radius ($r_{\rm s}$) away from their black hole (BH, Tombesi et al. 2012). Therefore, one generally believes that the UFOs may be the winds produced from the accretion disk around the center BH (e.g. Proga \& Kallman 2004; Fukumura et al. 2010; Nomura \& Ohsuga 2017; Yang et al. 2021).

The Fe K$\alpha$ UFOs are detected not only in a significant fraction ($>35\%$) of radio-quiet AGNs but also in a small sample of radio-loud AGNs (Tombesi et al. 2012; 2014). The UFOs in radio-loud AGNs are detected at the inclination angle of $\sim10^{\rm o}$--$70^{\rm o}$ away from jets (Tombesi et al. 2014). Generally, it is believed that ``winds'' associate with radio-quiet AGNs while ``jets'' associate with radio-loud AGNs. The discovery of UFOs in radio-loud AGNs challenges the simplistic wind-jet dichotomy (Fukumura et al. 2014).

The formation of Fe K$\alpha$ UFOs seems to be explained well in both a magnetic-driving model and a radiative-pressure (such as line force) driving model (Proga \& Kallman 2004; Fukumura et al. 2014; 2018; Nomura et al. 2016; Yang et al. 2021). Magnetic driving is proposed to drive UFOs (e.g. Fukumura et al. 2014; 2018). For example, Fukumura et al. (2018) have suggested that the UFOs observed in PDS 456 can be driven only by magnetic fields. It is noted that Fukumura et al.'s works (2014; 2018) are based on a self-similar solution. The self-similar solution is worth studying again based on numerical simulations.

When the mechanism of line-force driving is used to understand the properties of UFOs, one has the following problems. (1) Line-force driving becomes efficient only when the gas is weakly ionized. When the ionization parameter $\xi>100$ erg s$^{-1}$ cm, the line force becomes negligible. However, the ionization parameter of UFOs is much greater than 100 erg s$^{-1}$ cm, which implies that the line force seems to be inefficient in driving UFOs. (2) Numerical simulations have indicated that the line-force-driven winds have a narrow opening angle (e.g. Nomura et al. 2016; Yang et al. 2021). For example, Yang et al. (2021) found that the line-force-driven high-velocity winds can form at the angular range of 57$^{\rm o}$--77$^{\rm o}$ away from the rotational axis. When a weak poloidal magnetic field is included, the angular range can become slightly broader. However, numerical simulations imply that the line-force-driven winds seem to fail to describe the low-inclination UFOs detected in radio-loud AGNs. (3) Numerical simulations have found that the line-force-driven winds can reach $\sim10\%$ of the light speed ($c$) when the typical parameters of AGNs are employed (Nomura et al. 2016; Yang et al. 2021). However, in some AGNs, the UFO speed exceeds 0.2 $c$ (e.g. Tombesi et al. 2010; Gofford et al. 2015). (4) Numerical simulations have also found that the line-force-driving mechanism becomes inefficient in faint AGNs whose luminosity is less than $\sim10\%$ of Eddington luminosity ($L_{\rm Edd}$, Nomura et al. 2016). However, the UFOs are also observed in the faint Seyfert 2 galaxies, such as NGC 2992, whose luminosity is estimated to be 0.001--0.04 $L_{\rm Edd}$ (Marinucci et al. 2018). Therefore, the line-force-driving mechanism is facing challenges when it is considered as the only mechanism of driving UFOs.

Magnetic driving and line-force driving are often individually studied as mechanisms of driving UFOs. However, the two mechanisms could simultaneously operate in driving UFOs. Motivated by this reason, we investigate in this paper the winds driven by both line force and magnetic field and try to explain the formation of UFOs in radio-loud AGNs.

This paper is organized as follows: section 2 introduces our model and method; section 3 is devoted to the analysis of simulations, and section 4 summarizes our results.

\section{Model and Method} \label{sec:Model}

In this paper, we study the disk winds driven by both line force and magnetic field from a thin disk around a BH, based on numerical simulations. This work is the extension of Yang et al.'s work (2021), which studied the line-force-driven winds in the case of weak magnetic fields. Here, we extend Yang et al.'s work (2021) to the case of strong magnetic fields. Our models and numerical method are in most respect as described in Yang et al. 'work (2021). In the following, we only describe the key elements of our models.

To simulate the winds driven by both line force and magnetic field from a thin disk, we axis-symmetrically solve the equations of magnetohydrodynamic (MHD) in spherical polar coordinates ($r$,$\theta$,$\phi$). The $\theta=0$ axis corresponds to the rotational axis of the thin disk. The equations of MHD are given as follows:
\begin{equation}
\frac{d\rho}{dt}+\rho\nabla\cdot {\bf v}=0,
\label{cont}
\end{equation}

\begin{equation}
\rho\frac{d {\bf v}}{dt}=-\nabla P-\rho\nabla
\psi+\frac{1}{4 \pi}(\nabla \times \bf B)\times {\bf B}+\rho {\bf F}^{\rm rad},
\label{monentum}
\end{equation}

\begin{equation}
\rho\frac{d(e/\rho)}{dt}=-P\nabla\cdot {\bf v}+\rho \dot{E},
\label{energyequation}
\end{equation}
and
\begin{equation}
\frac{\partial {\bf B}}{\partial t}=\nabla\times({\bf v}\times{\bf B}),
\label{induction_equation}
\end{equation}
where $d/dt(\equiv \partial / \partial t+ \mathbf{v} \cdot \nabla)$ denotes the Lagrangian derivative. The dependent quantities $\rho$ , $P$, $\mathbf{v}$, and $e$ are gas density, gas pressure, velocity, and internal energy, respectively, $\bf{B}$ is the magnetic field and $\psi$ is the pseudo-Newtonian potential of a BH (Paczy\'{n}sky \& Wiita 1980). $\mathbf{F}^{\text{rad}}$ is the sum of Compton-scattering force and line force exerted on per unit mass. The calculation of $\mathbf{F}^{\text{rad}}$ is referred to Proga et al. (1998); $\rho\dot{E}$ is the net heating rate per unit mass, which is also referred to equations (18)--(21) in Proga et al. (2000). We adopt an equation of state of ideal gas $P\equiv(\gamma -1)e\equiv\rho \kappa T_{\rm g}/\mu m_{\rm p}$ and set $\gamma =5/3$, where $\mu$, $\kappa$, $m_{\text{p}}$, and $T_{\rm g}$ are the mean molecular weight, the Boltzmann constant, the proton mass, and the gas temperature, respectively.

\begin{table*}
\begin{center}

\caption{Summary of models}

\begin{tabular}{cccccccccc}
\hline\noalign{\smallskip} \hline\noalign{\smallskip}

Run & $\rho_{\rm d}$ (g/cm$^3$) & $\beta_{0}$ & $\alpha_{0}$   & $\epsilon$  & $f_{\rm X}$  & $\dot{M}_{\rm w}$  & $P_{\text{m,w}}$ & $P_{\text{k,w}}$   & $P_{\text{th,w}}$  \\
    &          &          &       &       &       &   ($M_{\odot}\cdot$yr$^{-1}$)  &  (g$\cdot$cm$\cdot$s$^{-2}$)  &   (erg$\cdot$s$^{-1}$)    &  (erg$\cdot$s$^{-1}$)   \\
(1) & (2)      & (3)      &  (4)  & (5)   & (6) & (7)  & (8) & (9) & (10)\\

\hline\noalign{\smallskip}

LDMHD0.3 & 10$^{-12}$ & $0.1$ &5.0 &0.3 & 8.96$\times10^{-2}$   & 5.48$\times$10$^{-2}$    &5.67$\times$10$^{33}$ & 6.38$\times$10$^{42}$ & 4.29$\times$10$^{41}$\\
LDMHD0.4 & 10$^{-12}$ & $0.1$ &5.0 &0.4 & 7.59$\times10^{-2}$  &  6.51$\times$10$^{-2}$    &7.97$\times$10$^{33}$ & 1.02$\times$10$^{43}$ & 4.16$\times$10$^{41}$ \\
LDMHD0.5 & 10$^{-12}$ & $0.1$ &5.0 &0.5 & 6.68$\times10^{-2}$   & 7.62$\times$10$^{-2}$    &9.66$\times$10$^{33}$ & 1.22$\times$10$^{43}$ & 4.72$\times$10$^{41}$ \\
LDMHD0.6 & 10$^{-12}$ & $0.1$ &5.0 &0.6 & 6.01$\times10^{-2}$   & 7.76$\times$10$^{-2}$    &8.42$\times$10$^{33}$ & 9.59$\times$10$^{42}$ & 4.07$\times$10$^{41}$ \\

LDMHD0.6a & 10$^{-12}$ & $0.2$ &5.0 &0.6 & 6.01$\times10^{-2}$  & 4.89$\times$10$^{-2}$    &5.25$\times$10$^{33}$ & 5.58$\times$10$^{42}$ & 1.62$\times$10$^{41}$ \\

LDMHD0.6b & 10$^{-12}$ & $100.0$ &5.0 &0.6 & 6.01$\times10^{-2}$   & 1.16$\times$10$^{-1}$     &1.12$\times$10$^{34}$ & 9.12$\times$10$^{42}$ & 5.81$\times$10$^{40}$ \\

LDMHD0.6c & 4$\times$10$^{-12}$ & $0.1$ &5.0 &0.6 & 6.01$\times10^{-2}$   & 9.13$\times$10$^{-2}$     &1.04$\times$10$^{34}$ & 1.35$\times$10$^{43}$ & 4.09$\times$10$^{42}$ \\

\hline\noalign{\smallskip} \hline\noalign{\smallskip}
\end{tabular}
\end{center}

\begin{list}{}
\item\scriptsize{Column 1: model names; column 2: the gas density at the disk surface; Columns 3 and 4: values of $\beta_0$ and $\alpha_0$, which determine the strength and inclination of initial magnetic fields, respectively; column 5: the ratio ($\epsilon$) of disk luminosity ($L_{\rm D}$) to Eddington luminosity ($L_{\rm Edd}$); column 6: the ratio of the X-ray luminosity to the disk luminosity; column 7: the time-averaged values of the mass outflow rate; columns 8--10: the time-averaged values of the momentum flux, kinetic energy flux, and thermal energy flux, respectively, which are carried out by the outflows at the outer boundary. }
\end{list}
\label{tab1e_1}
\end{table*}

We briefly describe the main parameters and setup of models in the following:

(1) The strength of line force is denoted by the force multiplier ($\mathcal{M}$), which is a function of the ionization parameter ($\xi$) and the local optical depth parameter (Rybichi \& Hummer 1978).  The ionization parameter $\xi$ is defined as $\xi=4\pi F_{_\text{X}}/n$, where $n$ is number density ($n=\rho/\mu m_{\text{p}}$) of the gas. We set the UV attenuation to be $0.4 \text{ g}^{-1}\text{cm}^2$ and the X-ray attenuation to be $0.4 \text{ g}^{-1}\text{cm}^2$ for $\xi\geq10^5$ erg s$^{-1}$ cm while $40 \text{ g}^{-1}\text{cm}^2$ for $\xi<10^5$ erg s$^{-1}$ cm; $\mu$ is set to be 1.0. The more details of calculating $\mathcal{M}$ are referred to equations (11)--(16) in Proga et al. (2000).

(2) The ratio ($f_{\text{X}}$) of the X-ray luminosity ($L_{\rm cor}$) to the disk luminosity ($L_{\rm D}$) is calculated using an observed nonlinear relation between the X-ray (2 keV) and UV (2500 {\AA}) emissions in quasars (Lusso \& Risaliti 2016). This relation implies that fewer X-ray photons (per unit UV luminosity) are emitted in optically bright AGNs than that in optically faint AGNs. The calculation of $f_{\rm X}$ is referred to Yang et al. (2021). The radiation temperature of X-ray emissions is set to be $10^8$ K.

(3) To satisfy $\nabla \cdot {\bf B}=0$, a vector potential ${\bf A}\equiv(0,0,{\bf A}_{\phi})^{\rm T}$ is used to set the initial configuration of magnetic fields. According to Cao \& Spruit (1994), the toroidal component ($
{\bf A}_{\phi}$) in spherical coordinates is given by
\begin{equation}
\begin{aligned}
{\bf A}_\phi=\frac{\Phi_{0}}{rsin(\theta)}\frac{r_{\text{in}}^2}{\alpha_0^2}[\sqrt{(\alpha_0\frac{r}{r_{\text{in}}}\sin(\theta))^2+(1+\alpha_0\frac{r}{r_{\text{in}}}\cos(\theta))^2}\\
-|1+\alpha_0\frac{r}{r_{\text{in}}}\cos(\theta)|],
\label{vectorpotential}
\end{aligned}
\end{equation}
where $r_{\text{in}}$ is the inner boundary of computational domain and $\Phi_{0}$ and $\alpha_0$ are the two parameters used to determine the strength and inclination of initial magnetic field, respectively. $\Phi_{0}^2$ is defined as $\Phi_{0}^2=P_{_{\rm s,}r_{\text{in}}}[1+\alpha_0^2+\sqrt{1+\alpha_0^2}]/\beta_0$, where $P_{_{\rm s,}r_{\text{in}}}$ is the gas pressure at the disk surface at $r=r_{\text{in}}$ and $\beta_{0}=(8\pi P_{_{\text{s},}r_{\text{in}}})/B({r_{\text{in}},\pi/2})^2$ is used to scale the magnitude of the magnetic field at $r=r_{\text{in}}$, respectively. Therefore, the poloidal field components (${\bf B}_{\rm p}\equiv{B}_r {\bf e}_r+{B}_\theta {\bf e}_\theta$) in the space above the disk are give by $B_{r}=\frac{1}{r sin(\theta)}\frac{\partial({\bf A}_\phi sin(\theta))}{\partial \theta}$ and $B_{\theta}=-\frac{1}{r}\frac{\partial({\bf A}_\phi r)}{\partial r}$, respectively. The magnetic field configuration is similar to that proposed in the self-similar disk winds by Blandford \& Payne (1982).

\begin{figure}
\includegraphics[width=.45\textwidth]{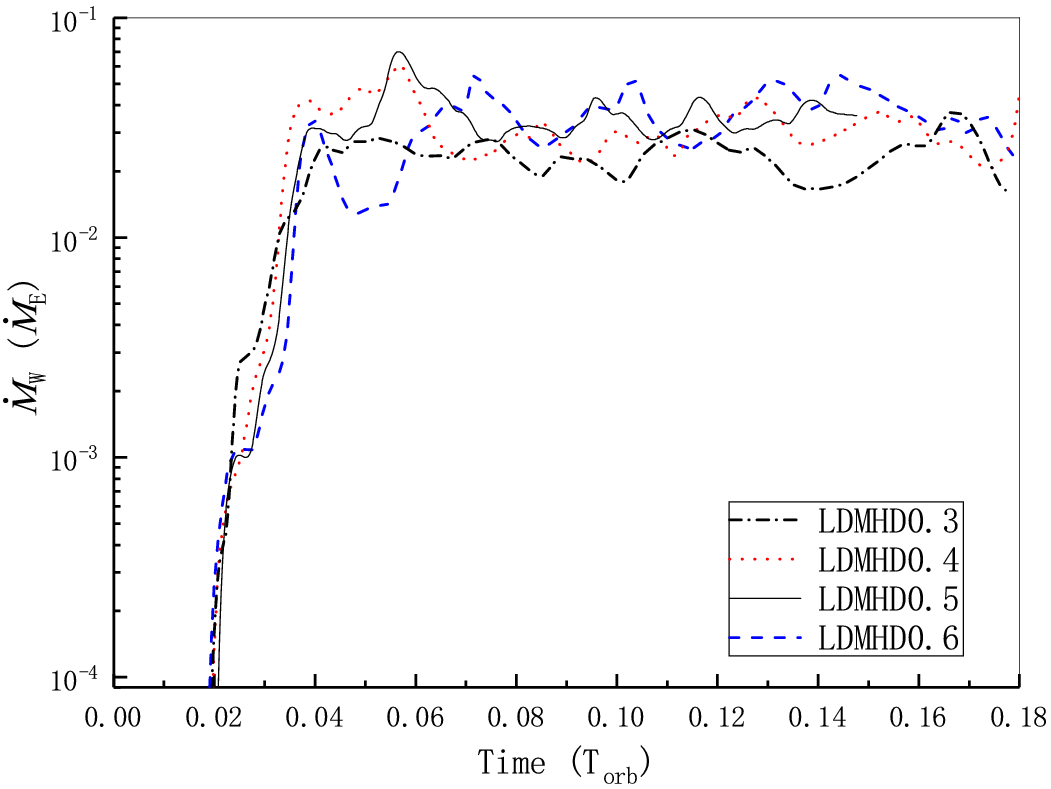}

\ \centering \caption{Time evolution of mass outflow rates ($\dot{M}_{\rm W}$) at the outer boundary for runs LDMHD0.3--LDMHD0.6. In this figure, the mass outflow rate is given in units of Eddington accretion rate ($\dot{M}_{\rm E}=10 L_{\rm Edd}/c^2$).}
\label{fig1}
\end{figure}

(4) Initially, we set the gas in isothermal equilibrium and force balance. We apply the outflow boundary condition at the inner and outer radial boundaries and the axially symmetric boundary condition at the pole (i.e. $\theta=0$), respectively. At the disk surface (i.e. $\theta=90^{\rm o}$), we fix the gas density and temperature to be $\rho_{\rm d}=10^{-12}$ g cm$^{-3}$ and $T_{\rm eff}(r_{\rm d})=(\pi I_{\rm d}(r_{\rm d})/\sigma)^{\frac{1}{4}}$, respectively, where $I_{\rm d}(r_{\rm d})$ is local isotropic intensity at the radial position of $r_{\rm d}$ on the disk surface. We also set both radial and vertical velocities to be null at the disk surface. In our models, the magnetic fields are stronger than the gas pressure at the disk surface. When the effect of magnetic fields is taken into account, the thin disk surface is not of Keplerian rotation. Cao (2012) gave the rotational angular velocity of the disk surface when a large-scale magnetic field threads the thin disk. Following Cao (2012), the rotational velocity of the disk surface is given by
\begin{equation}
{\bf v}_{\phi}(r,\pi/2)^2=(r\Omega_{\rm K})^2 (1-\frac{2\beta \tilde{H}_{_{\rm D}}}{\kappa_{0}(1+\tilde{H}_{_{\rm D}}^2)^{3/2}}),
\label{rotating_velocity}
\end{equation}
where $\beta$ equals $B_{\rm z}(\pi/2)^2/2 P_{\rm s}$ ($P_{\rm s}$ is the gas pressure at the disk surface), and $\Omega_{\rm K}$, $\kappa_{0}$, and $\tilde{H}_{_{\rm D}}$ are Keplerian angular velocity of the disk surface at radius $r$, the slope of magnetic field lines at the disk surface, and the ratio of disk half thickness to disk radius, respectively. We set $\tilde{H}_{_{\rm D}}=0.01$ here. For the magnetic fields, the reflected boundary condition is used to the plane of $\theta=90^{\rm o}$.

(5)The computational domain is divided into 144$\times$160 zones. In the $r$ direction, there are 144 zones and the radial size ratio is set to be $(\bigtriangleup r)_{i+1} / (\bigtriangleup r)_{i} = 1.04$. In the $\theta$ direction, 16 zones are uniformly distributed over the angular range of $0^{\rm o}$--$15^{\rm o}$ while 144 zones are non-uniformly distributed over the angular range of $15^{\rm o}$--$90^{\rm o}$. The angular size ratio is set to be $(\bigtriangleup \theta)_{j+1} / (\bigtriangleup \theta)_{j} = 0.970072$.

\begin{figure*}
\includegraphics[width=.32\textwidth]{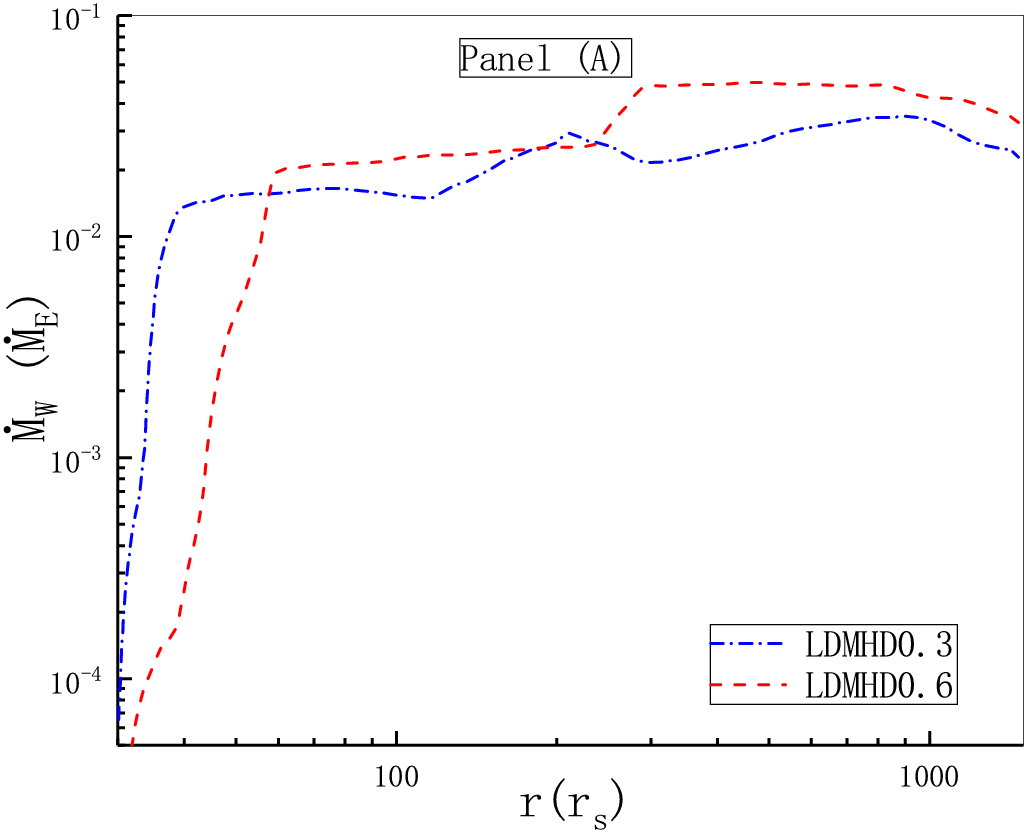}
\includegraphics[width=.32\textwidth]{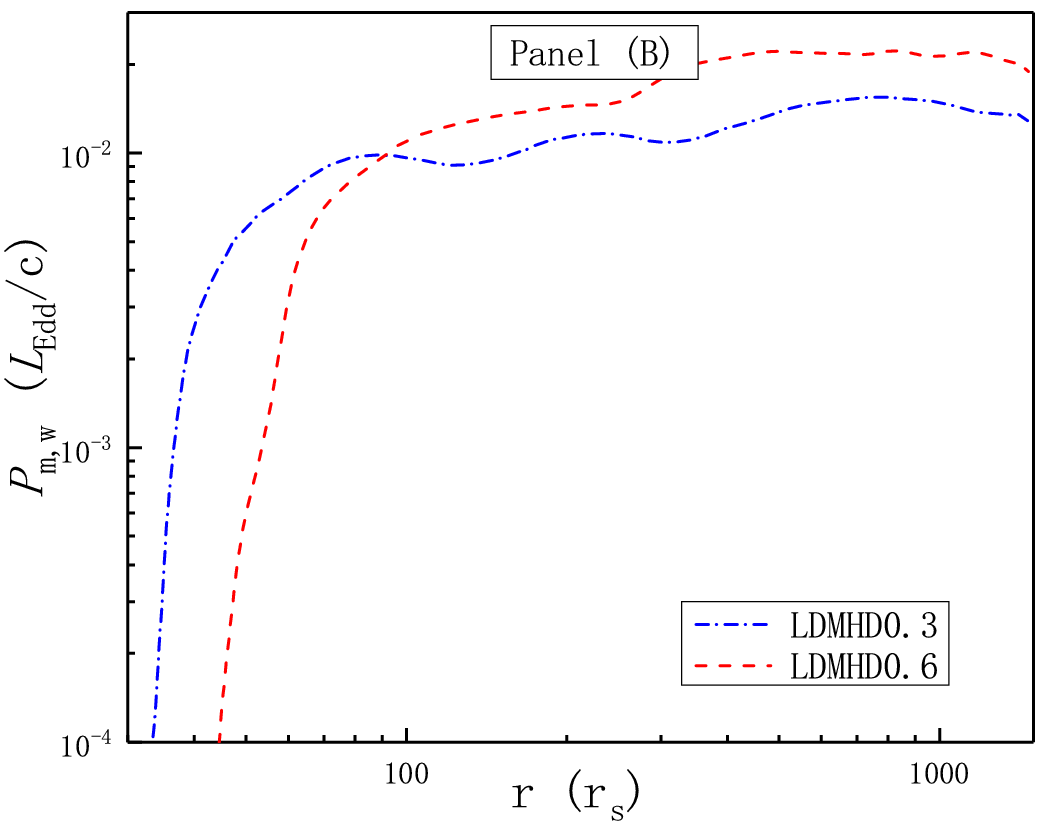}
\includegraphics[width=.32\textwidth]{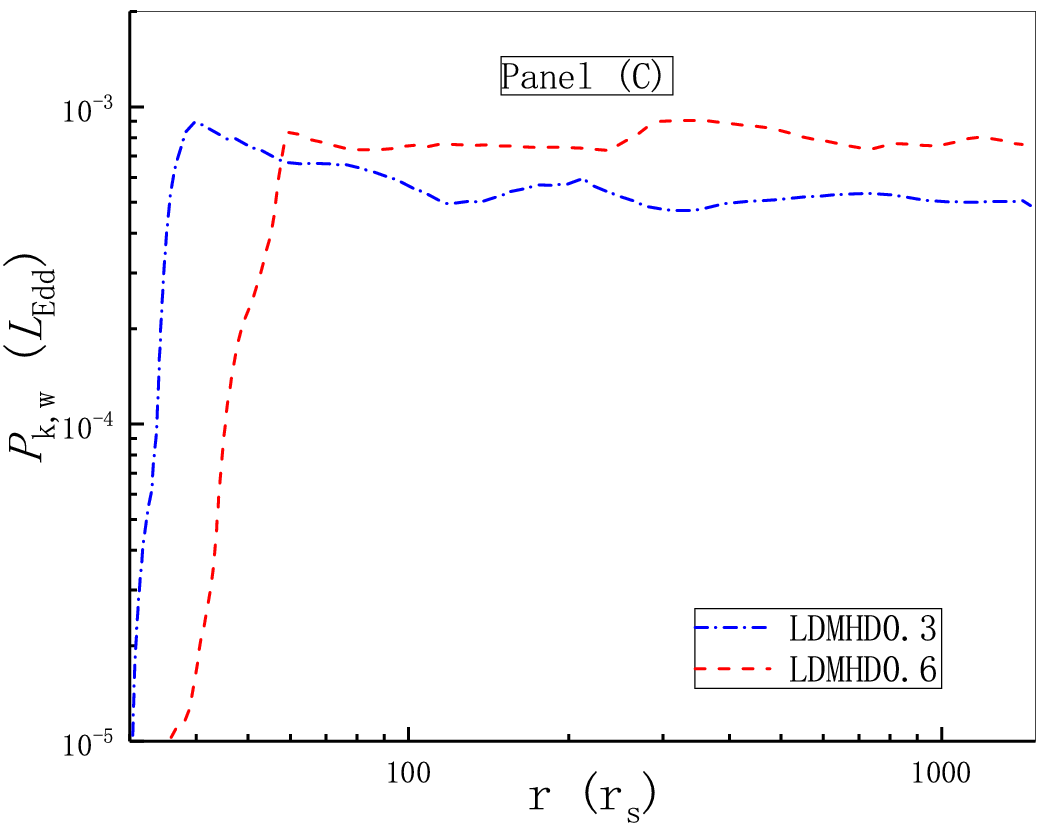}

\ \centering \caption{Radial dependence of time-averaged quantities. Panels A--C shows the mass outflow rate, the momentum flux of winds, and the kinetic power of winds, respectively. In panel A, $\dot{M}_{\rm E}=10 L_{\rm Edd}/c^2$.}
\label{fig2}
\end{figure*}

\begin{figure*}
\includegraphics[width=.32\textwidth]{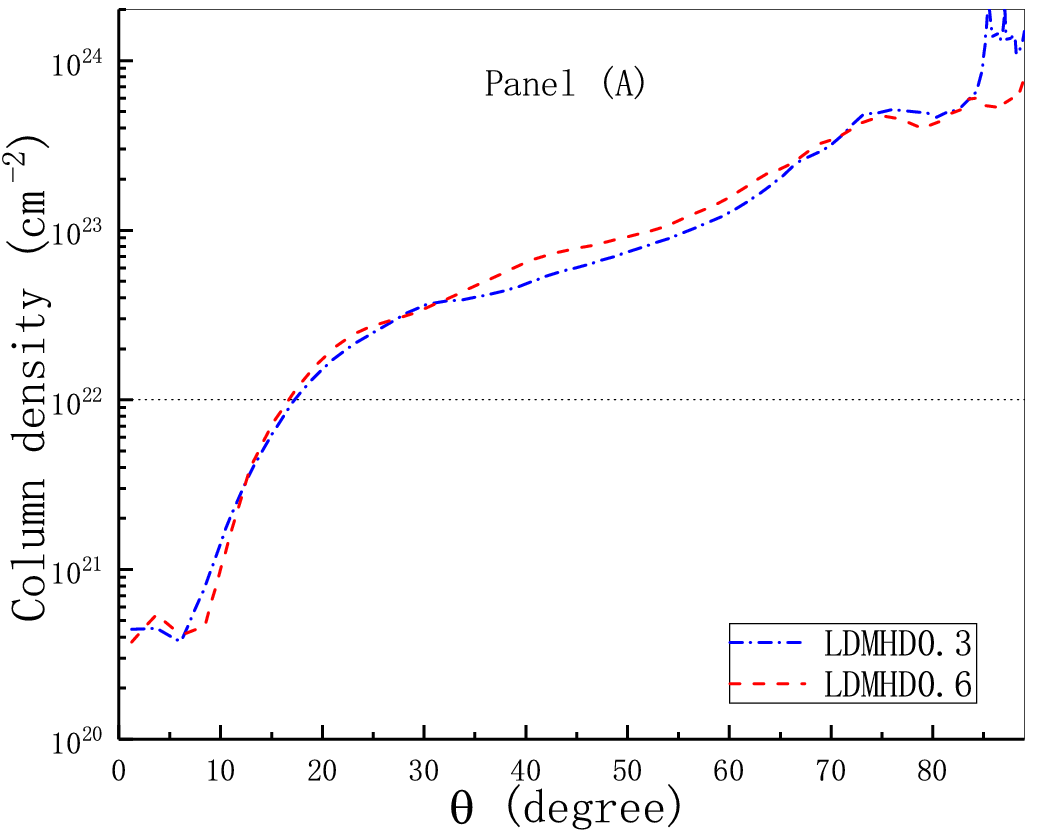}
\includegraphics[width=.32\textwidth]{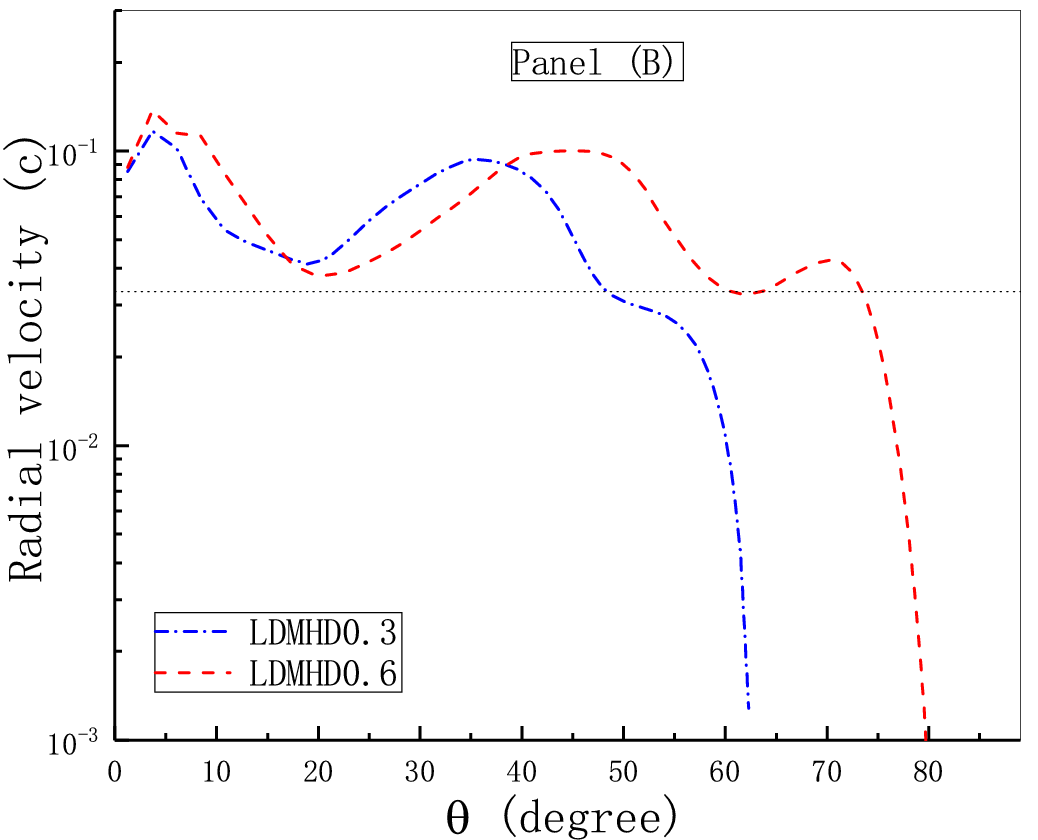}
\includegraphics[width=.32\textwidth]{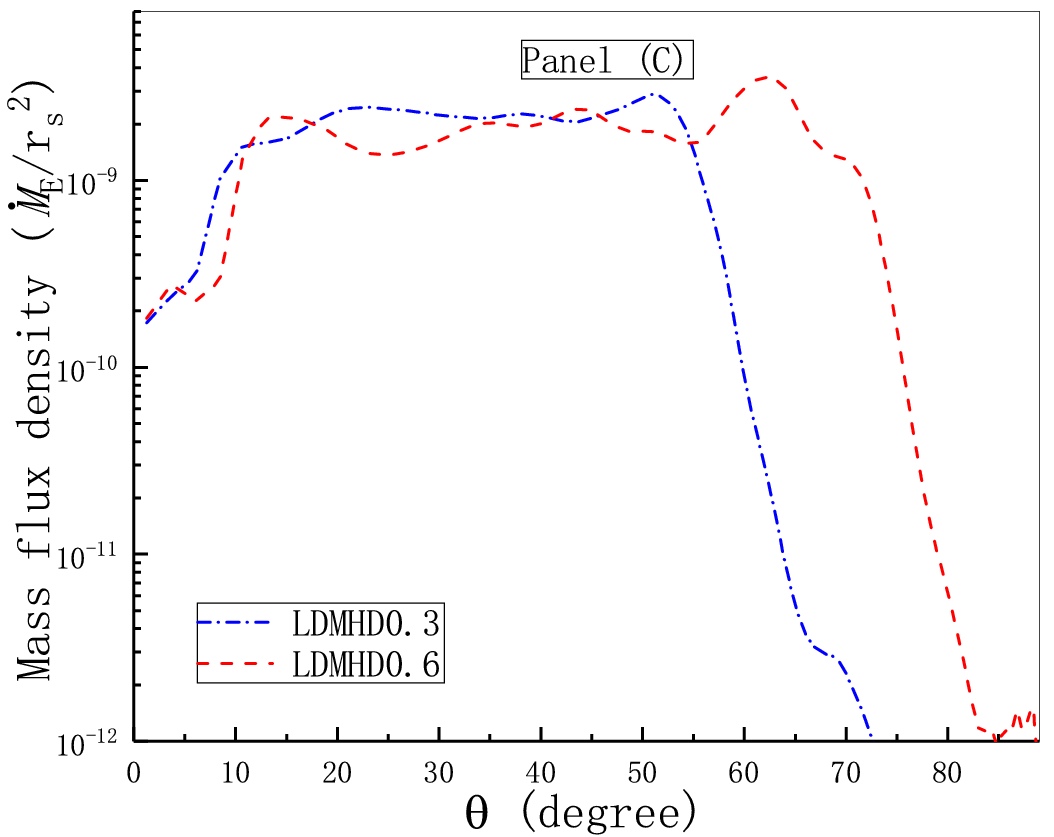}

\ \centering \caption{Angular profiles of a variety of time-averaged variables. Panels A--C give the angular profiles of the column density, the radial velocity at the outer boundary, and the mass flux density at the outer boundary, respectively. The dotted line in panel A means the column density of $10^{22}$ cm$^{-2}$. The dotted line in panel B means the velocity of $10^4$ km s$^{-1}$.}
\label{fig3}
\end{figure*}

\section{Results} \label{sec:Results}

\subsection{Basic properties of winds} \label{subsec:Basic properties}

When a magnetic field is strong, simulating cannot last for too long because the time step becomes too short. The solutions are considered to reach a quasi-steady state when the change of solutions is not significant within 0.05 $T_{\rm orb}$, where $T_{\rm orb}$ is the Keplerian orbital period at the outer boundary. We have implemented simulations of seven models for testing the effects of the disk luminosity and the magnetic field strength. The seven models can quickly reach a quasi-steady state. They are summarized in Table 1, where columns 2--6 give the main parameters of models while columns 8--11 give the time-averaged quantities of winds at the outer boundary, such as the mass outflow rate ($\dot{M}_{\rm w}$), the momentum flux ($P_{\rm w}$), the kinetic ($P_{\rm k,w}$) energy flux, and the thermal energy ($P_{\rm th,w}$) fluxes, respectively. Runs LDMHD0.3--LDMHD0.6 are used to test the effects of disk luminosity while runs LDMHD0.6a and LDMHD0.6b are employed to test the effects of initial magnetic field strength. The effects of magnetic field strength will be discussed in Section 3.4. Figure 1 shows time evolution of mass outflow rate ($\dot{M}_{\rm W}$) at the outer boundary for runs LDMHD0.3--LDMHD0.6. The disk luminosity does not significantly affect the mass outflow rate of winds, as shown by Figure 1. On average, the higher disk luminosity is helpful to enhance the mass outflow rate. For example, runs LDMHD0.5 and LDMHD0.6 have higher mass outflow rates than LDMHD0.3 and LDMHD0.4 on average. For run LDMHD0.3--LDMHD0.6, there is no order-of-magnitude difference in their luminosity, so that their mass outflow rates do not differ much. According to Figure \ref{fig1}, time-averaging is implemented over the time interval of 0.089--0.127 $T_{\rm orb}$, including 250 data files. The time interval is equivalent to 13.4 orbit periods at the inner boundary and much longer than the dynamical scale of winds from the inner boundary to the outer boundary. The averaged quantities thereafter are also calculated in this way, expect for the pure MHD models in section 3.5.

To describe the radial dependence of outflow properties, we further calculate the radial dependence of the mass outflow rate ($\dot {M}_{\rm w} (r)$), the momentum flux ($P_{\rm m,w} (r)$) of winds, and the kinetic power ($P_{\rm k,w} (r)$) of winds. According to Table 1, the thermal energy flux of winds is much less than the kinetic energy flux of the winds. Therefore, compared to the kinetic energy of winds, the thermal energy of winds is not important so that it is not necessary to investigate the radial dependence of the thermal energy carried out by winds. $\dot {M}_{\rm w} (r)$, $P_{\rm k,w} (r)$, and $P_{\rm m,w} (r)$ are, respectively, given by
\begin{equation}
\dot {M}_{\rm w} (r)=4\pi r^2 \int_{\rm 0^\circ}^{\rm 89^\circ}
\rho \max (v_r, 0) \sin\theta d\theta,
\end{equation}

\begin{equation}
P_{\rm m,w} (r)=4\pi r^2 \int_{\rm 0^\circ}^{\rm 89^\circ} \rho
\max(v^2_r,0) \sin\theta d\theta,
\end{equation}

and

\begin{equation}
P_{\rm k,w} (r)=2\pi r^2 \int_{\rm 0^\circ}^{\rm 89^\circ} \rho
\max(v_r^3,0) \sin\theta d\theta.
\end{equation}
Figure \ref{fig2} shows the radial dependence of the time-averaged $\dot {M}_{\rm w} (r)$, $P_{\rm m,w} (r)$, and $P_{\rm k,w} (r)$. As shown in Figure \ref{fig2}, for run LDMHD0.6, $\dot{M}_{\rm w}$, $P_{\rm k,w} (r)$, and $P_{\rm m,w} (r)$ rapidly increase inside $\sim65$ r$_{\rm s}$, and then $\dot{M}_{\rm w}$ and $P_{\rm m,w} (r)$ slightly increase outside $\sim65$ $\rm{r}_{\rm s}$ while $P_{\rm k,w} (r)$ almost keeps constant outside $\sim65$ $\rm{r}_{\rm s}$. For run LDMHD0.3, the radial dependence of $\dot {M}_{\rm w} (r)$, $P_{\rm m,w} (r)$, and $P_{\rm k,w} (r)$ is similar to the case in run LDMHD0.6. Table 1 and Figure 1 also imply that the mass outflow rate and the kinetic and thermal energy fluxes in run LDMHD0.6 are slightly higher than that that in run LDMHD0.3.

Figure \ref{fig3} shows the angular dependence of wind properties at the outer boundary. Panels A--C in Figure \ref{fig3} show the angular distribution of the column density ($N_{H}=\int^{1500r_{\rm s}}_{30r_{\rm s}}\frac{\rho(r,\theta)}{\mu m_{p}} dr$), the radial velocity, and the mass flux density, respectively. As shown in Figure \ref{fig3}, the column density is higher than $10^{22}$ cm$^{-2}$ at $\theta > \sim16^{\rm o}$ for run LDMHD0.3 and run LDMHD0.6. For run LDMHD0.6, the radial velocity of winds is higher than 10$^4$ km$\cdot$s$^{-1}$ at $\theta<73^{\rm o}$  and the maximum radial velocity of dense winds is about 0.1 $c$ at $\sim45^{\rm o}$. For run LDMHD0.3, the wind radial velocity at $\theta<48^{\rm o}$ is higher than 10$^4$ km s$^{-1}$ and the maximum radial velocity of dense winds is about 0.094 $c$ at $\sim35^{\rm o}$. The mass flux density of winds is comparable over $\theta<55^{\rm o}$ in runs LDMHD0.3 and LDMHD0.6. We define the high-velocity winds as the outflows that have a velocity higher than 10$^4$ km s$^{-1}$ and the column density higher than $10^{22}$ cm$^{-2}$. Comparing runs LDMHD0.3 and LDMHD0.6, the high-velocity winds of run LDMHD0.6 have a slightly broader opening angle and a slightly higher maximum speed.

\begin{figure}
\includegraphics[width=.45\textwidth]{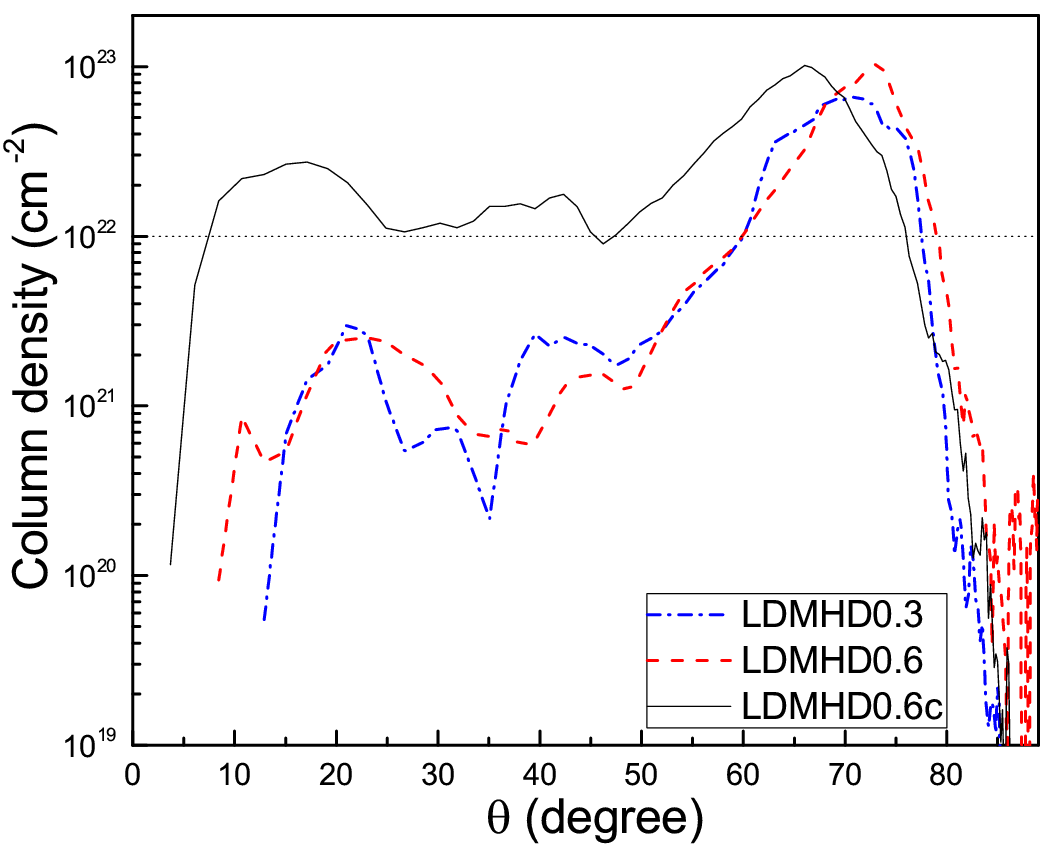}

\ \centering \caption{Angular profile of column density for the matter with $3\leq log(\xi/(\rm{erg s}^{-1}\rm{ cm}))\leq6$ and $v_{\rm r}\geq 10^4\rm{km} \cdot \rm{s}^{-1}$ for runs LDMHD0.3, LDMHD0.6, and LDMHD0.6c.}
\label{fig4}
\end{figure}

UFOs are detected at hard X-ray bands and often used to define the absorber, whose velocity is higher than $10^4\rm{ km}\cdot \rm{s}^{-1}$ and whose ionization parameter ($\xi$) and column density are distributed in the range of  log($\xi$/(erg s$^{-1}$ cm))$\sim$3--6 and  $10^{22}$ cm$^{-2}$$\lesssim N_{\rm H}\lesssim$ $10^{24}$ cm$^{-2}$, respectively (Tombesi et al. 2010). To investigate whether UFO could be detected from our models or not, we pick out the matter with $3\leq log(\xi/(\rm{erg s}^{-1}\rm{ cm}))\leq6$ and $v_{\rm r}\geq 10^4\rm{km} \cdot \rm{s}^{-1}$ from each snapshot and then calculate the column density of the matter. Finally, the column density is time averaged. Figure \ref{fig4} shows the angular distribution of column density of the matter with $3\leq log(\xi/(\rm{erg s}^{-1}\rm{ cm}))\leq6$ and $v_{\rm r}\geq 10^4\rm{km} \cdot \rm{s}^{-1}$. As shown in Figure \ref{fig4}, for runs LDMHD0.3 and LDMHD0.6, the column density is higher than $10^{22}$ cm$^{-2}$ over 60$^{\rm o}$--79$^{\rm o}$, where UFOs could form. For run LDMHD0.6c, the angular range of the column density higher than $10^{22}$ cm$^{-2}$ is 8$^{\rm o}$--77$^{\rm o}$. We conjecture that UFOs could be formed on the angular range of $\sim$8$^{\rm o}$--76$^{\rm o}$ for run LDMHD0.6c. UFOs could be detected in all of the runs in Table 1. Higher disk surface density is helpful to UFOs to be formed on a wider angle.

Tombesi et al. (2014) have searched for the UFOs in a heterogeneous sample of 26 radio-loud AGNs and suggested some potential correlations between the inclination angle and the UFO properties, such as the column density, the ionization parameter, and the outflow velocity (see their Figure 5). Their Figure 5 shows that most of the objects at the low-inclination angle have the relatively low outflow velocity, the relatively high ionization parameter, and the relatively low column density. In run LDMHD0.6c, compared with the low-latitude region (i.e. the region with high-inclination angle), at the high-latitude region (i.e. the region with low-inclination angle), the column density is relatively low (see Figure 4), and the ionization parameter and the UFO velocity is relatively high. The statistical confidence of Tombesi et al.'s result (2014) is not high, because their Figure 5 gives the distribution of seven objects.

The called non-canonical UFOs are detected at soft X-ray bands (Pounds et al. 2016; Serafinelli et al. 2019; Reeves et al. 2020). Both their column density and ionization parameter are low and distributed in the range of $N_{H}\sim10^{20{\rm-}22}$ cm$^{-2}$ and log($\xi$/(erg s$^{-1}$ cm))$\sim$0--3, respectively. The non-canonical UFOs have the properties different from the Fe K UFOs (Serafinelli et al. 2019). According to our simulations, we cannot detect the non-canonical UFOs that have low column density and low ionization parameter. Reeves et al. (2020) pointed out the soft X-ray UFOs to be located on parsec scales rather than around the accretion disk.

\subsection{Two-dimensional properties of winds} \label{subsec:2D structure}

\begin{figure*}
\includegraphics[width=.32\textwidth]{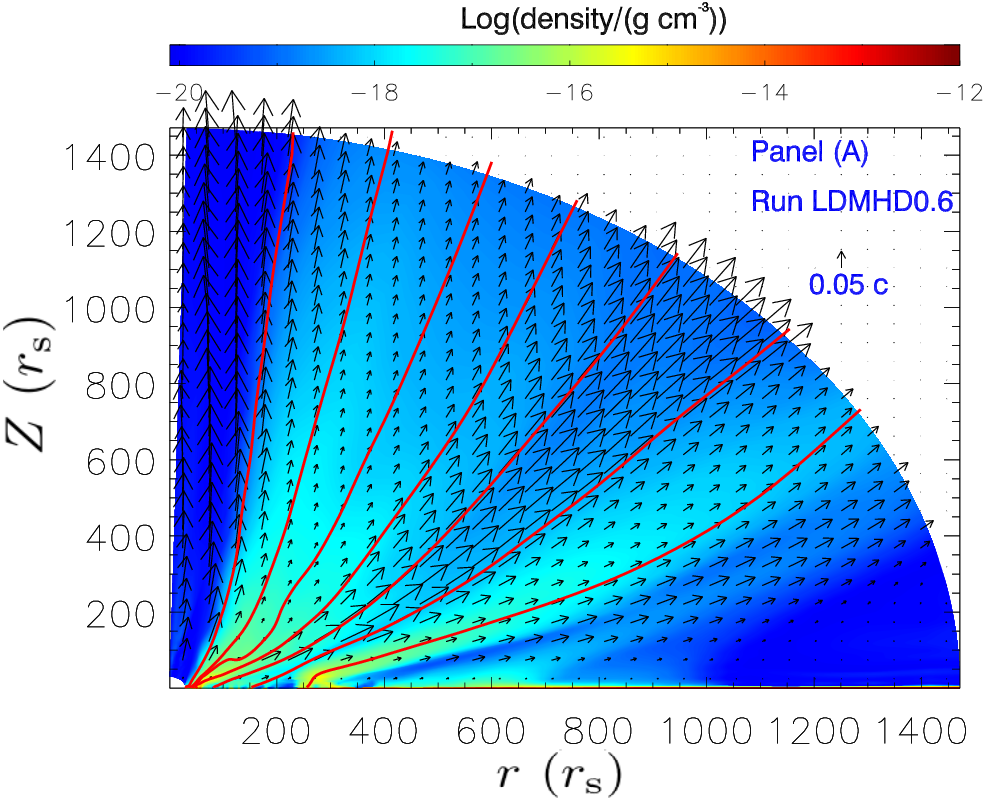}
\includegraphics[width=.32\textwidth]{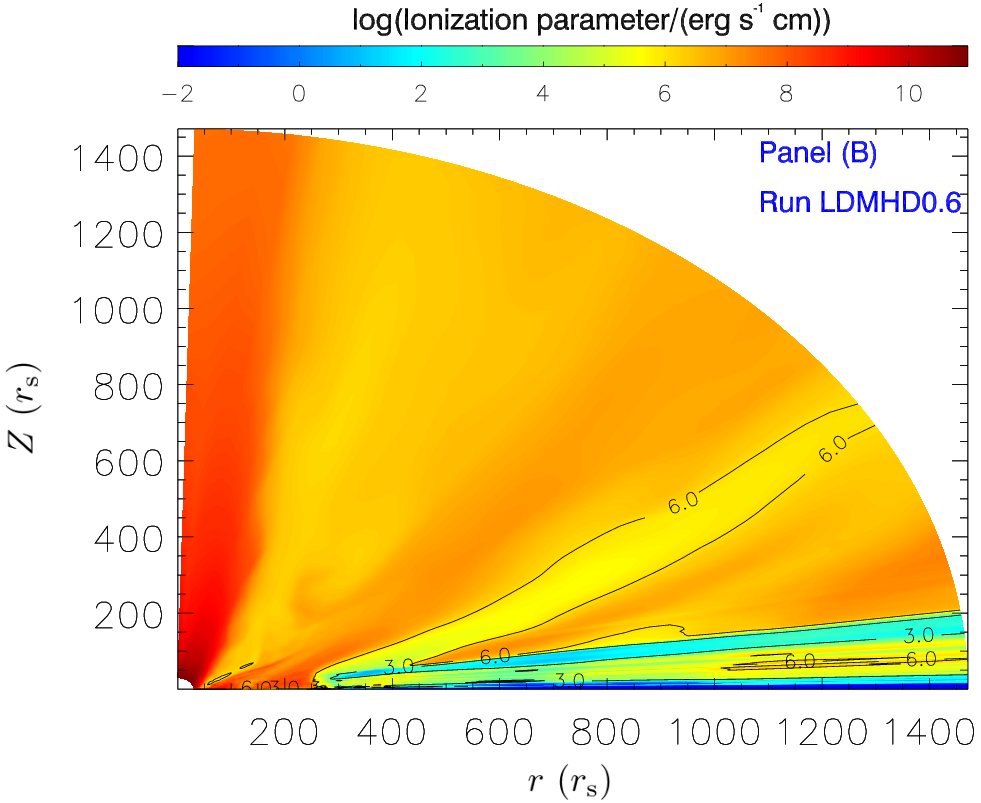}
\includegraphics[width=.32\textwidth]{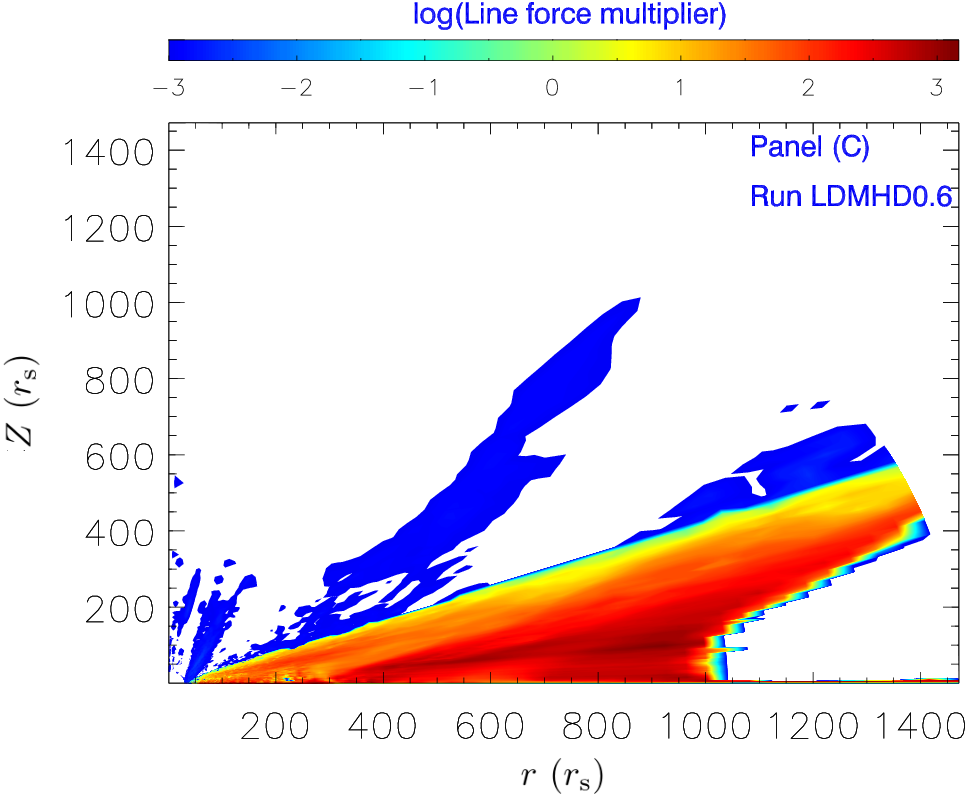}

\ \centering \caption{Two-dimensional structure of winds for run LDMHD0.6. (A): density distribution (color), poloidal velocity (arrows), and streamlines (red solid lines); (B): the ionization parameter ($\xi$); (C): the line force multiplier. }
\label{fig5}
\end{figure*}

\begin{figure*}
\includegraphics[width=.32\textwidth]{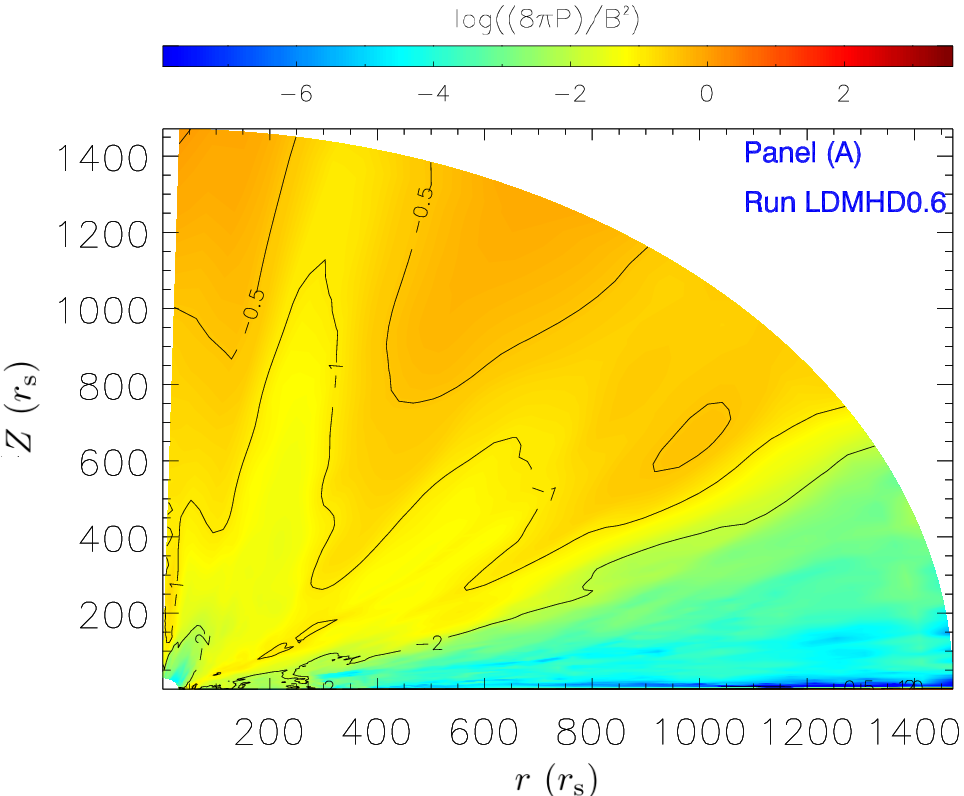}
\includegraphics[width=.32\textwidth]{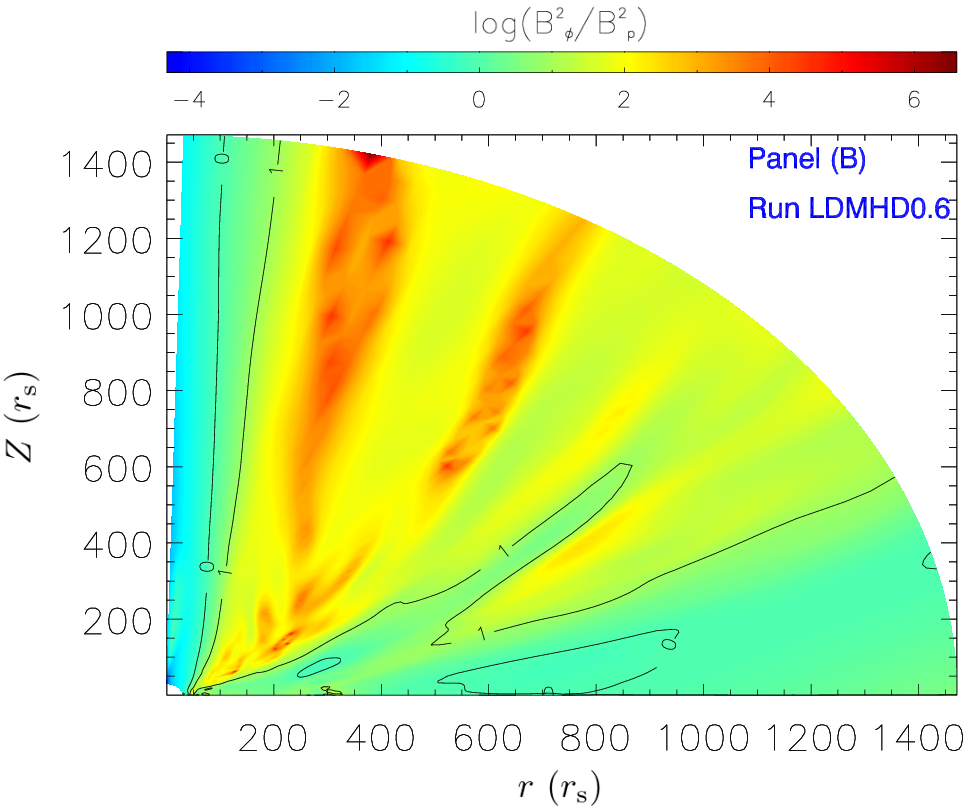}
\includegraphics[width=.32\textwidth]{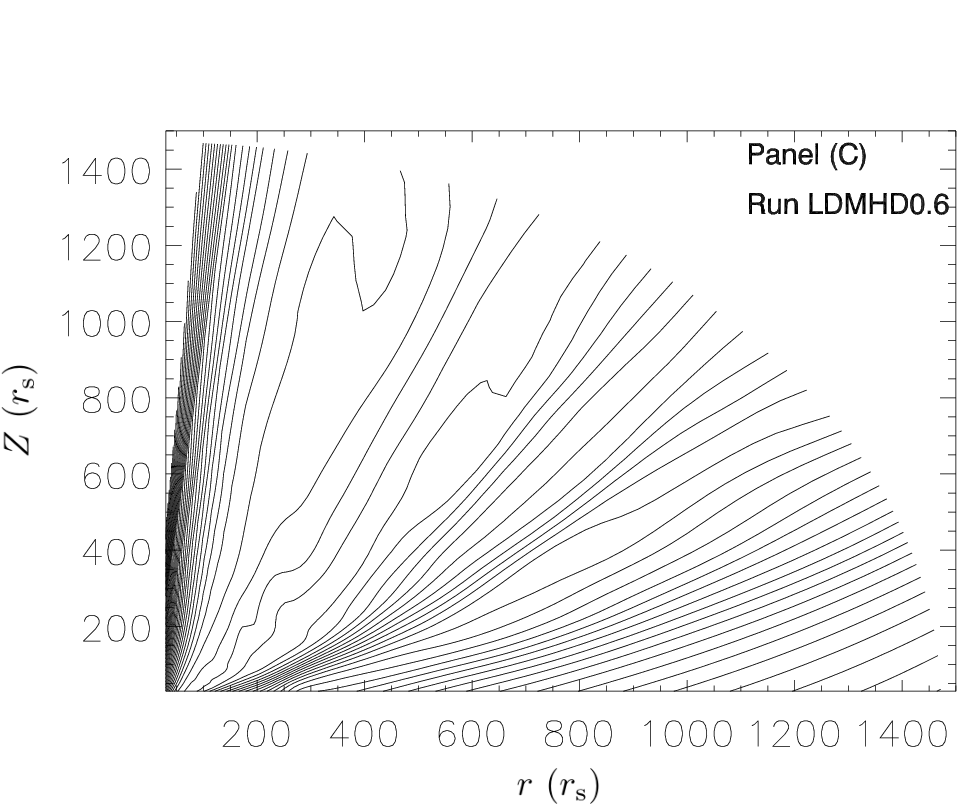}

\ \centering \caption{Two-dimensional structure of magnetic fields for run LDMHD0.6. (A): the logarithm of the ratio of gas pressure ($P$) to magnetic pressure ($B^2/8\pi$); (B) shows the logarithm of the ratio ($B^2_{\phi}/B^2_{\rm p}$) of toroidal ($B^2_{\phi}/8\pi$) to poloidal ($B^2_{\rm p}/8\pi$) magnetic pressure; panel C: the magnetic field lines on the $r$--$z$ plane. }
\label{fig6}
\end{figure*}

We take run LDMHD0.6 as an example to show the time-averaged two-dimensional structure of winds and magnetic field in Figures \ref{fig5} and \ref{fig6}. As shown in panel A of Figure \ref{fig5}, most of the materials were blown away from the disk surface inside 300$r_{\rm s}$. Panel B shows that the moderately ionized gases (log($\xi$/(erg s$^{-1}$ cm))$\sim$3--6) is mainly distributed over $\theta=\sim60^{\rm o}$--$90^{\rm o}$ in the sense of time average. In snapshots, the moderately ionized gases are clumpy and their distribution changes with time over $\theta=\sim10^{\rm o}$--$70^{\rm o}$ . When time-averaging is implemented, the clumpy gases with a moderate ionization degree are smoothed, and then the gases over $\theta=\sim10^{\rm o}$--$60^{\rm o}$ seem to be highly ionized, as shown in panel B. Figure \ref{fig4} implies that the clumpy gases with a moderate ionization degree could contribute to the observed features of UFOs. Panel C shows that line force becomes available at $\theta>70^{\rm o}$ (also see Figures 7 and 8). At the large radius, line force becomes invalid around the equator.

The line-force multiplier depends on the ionization degree (i.e. $\xi$) of gases. The ionization degree is determined by the gas density and the number of local X-ray photons. When $\xi$ increases from 0 to $\sim3$ erg s$^{-1}$ cm, the maximum value of $\mathcal{M}$ increases gradually from $\sim2000$ to $5000$ (Proga et al. 2000). When $\xi\gtrsim 10^2$ erg s$^{-1}$ cm, line force can be neglected because the maximum value of $\mathcal{M}$ is close to zero. This implies that the line force is effective in driving the low-ionized gases while ineffective in driving the moderately ionized gases. As in panel B of Figure \ref{fig5}, the gases at $\theta>\sim60^{\rm o}$ are moderately or highly ionized in the sense of time average. Over the angle range, however, the line force is also effective, as in panel C of Figure \ref{fig5}. The reasons are as follows. The distribution of dense gases in the inner region changes with time. When the X-ray photons are shielded by the inner-region dense gases at $\theta>\sim70^{\rm o}$, the moderately ionized winds become so lowly ionized that line-force driving becomes effective. For example, Panel B of Figure 8 also shows that line force is stronger than the BH gravitation force over $73^{\rm o}<\theta<85^{\rm o}$ in the sense of time average. When the low ionization winds at $\theta>\sim70^{\rm o}$ are in exposure to the X-ray again, they become moderately or highly ionized. Therefore, the winds at $\theta>\sim70^{\rm o}$ are intermittently accelerated by line force to high speed and they are moderately or highly ionized in the sense of time average.

For run LDMHD0.3, line force operates at angles that are narrower than that in run LDMHD0.6. Outflows in run LDMHD0.3 carry away lower kinetic energy than run LDMHD0.6 due to the low luminosity of run LDMHD0.3 (see panel C of Figure 2). In run LDMHD0.3, the low-latitude outflows ($\theta>60^{\rm o}$) from the inner region move to the large radius and then diffuse into the higher-latitude region due to the effect of magnetic fields. From panels B and C in Figure 3, it is seen that outflows are barely detected at the outer boundary when $\theta>60^{\rm o}$, for run LDMHD0.3. Panel A in Figure 3 shows that runs LDMHD0.3 and LDMHD0.6 have similar distribution for column density, i.e. the effect of ``X-ray  shielding'' is similar in runs LDMHD0.3 and LDMHD0.6. However, in run LDMHD0.3, the gas density is low in the low-latitude ($\theta>60^{\rm o}$) region outside $\sim 1000r_{\rm s}$, so that the gases have a high ionization degree. Therefore, line force in run LDMHD0.3 operates inside $\sim 1000r_{\rm s}$ while it is almost invalid outside $\sim 1000r_{\rm s}$.

\begin{figure}
\includegraphics[width=.45\textwidth]{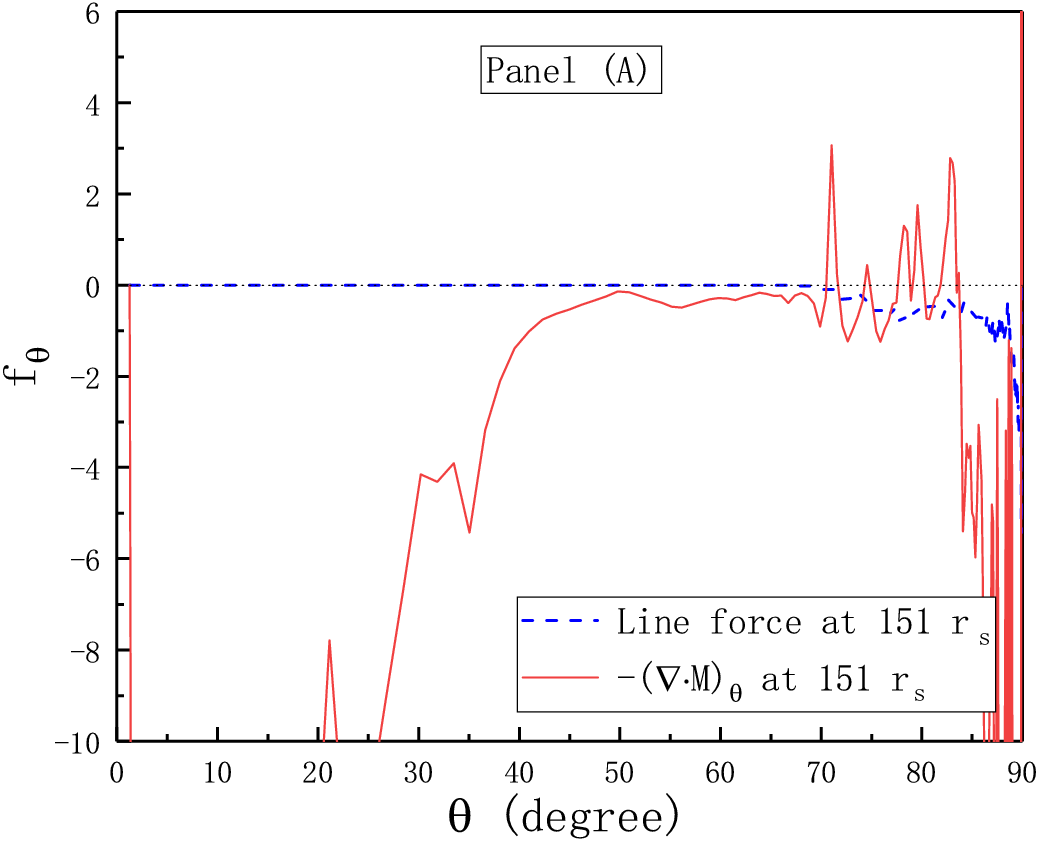}
\includegraphics[width=.45\textwidth]{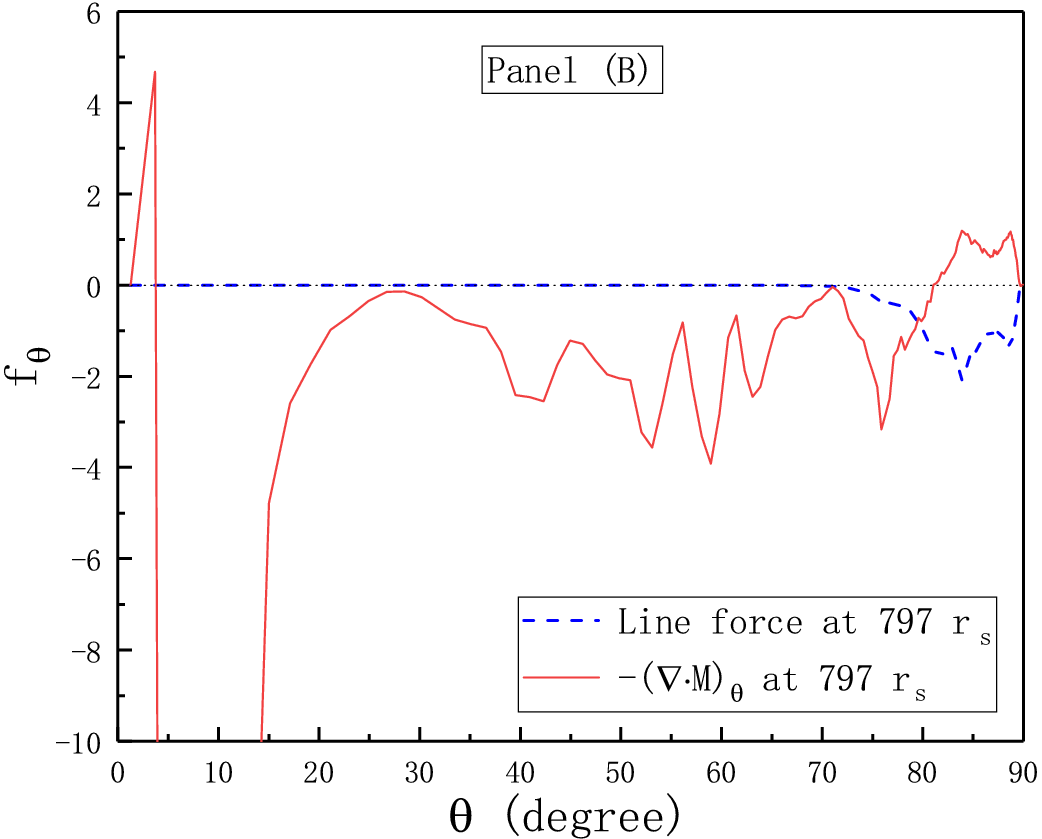}

\ \centering \caption{Angular dependence of angular forces in units of BH gravitation force, for run LDMHD0.6. In this figure, dotted-dashed lines means line force; solid lines mean $-(\nabla\cdot \bf{M})_{\rm \theta}$, where $\bf{M}$ is the magnetic stress tensor. Panels A and B show the angular components of line force and Lonertz force at 151 and 797 $r_{\rm s}$, respectively.}
\label{fig7}
\end{figure}

\begin{figure}
\includegraphics[width=.45\textwidth]{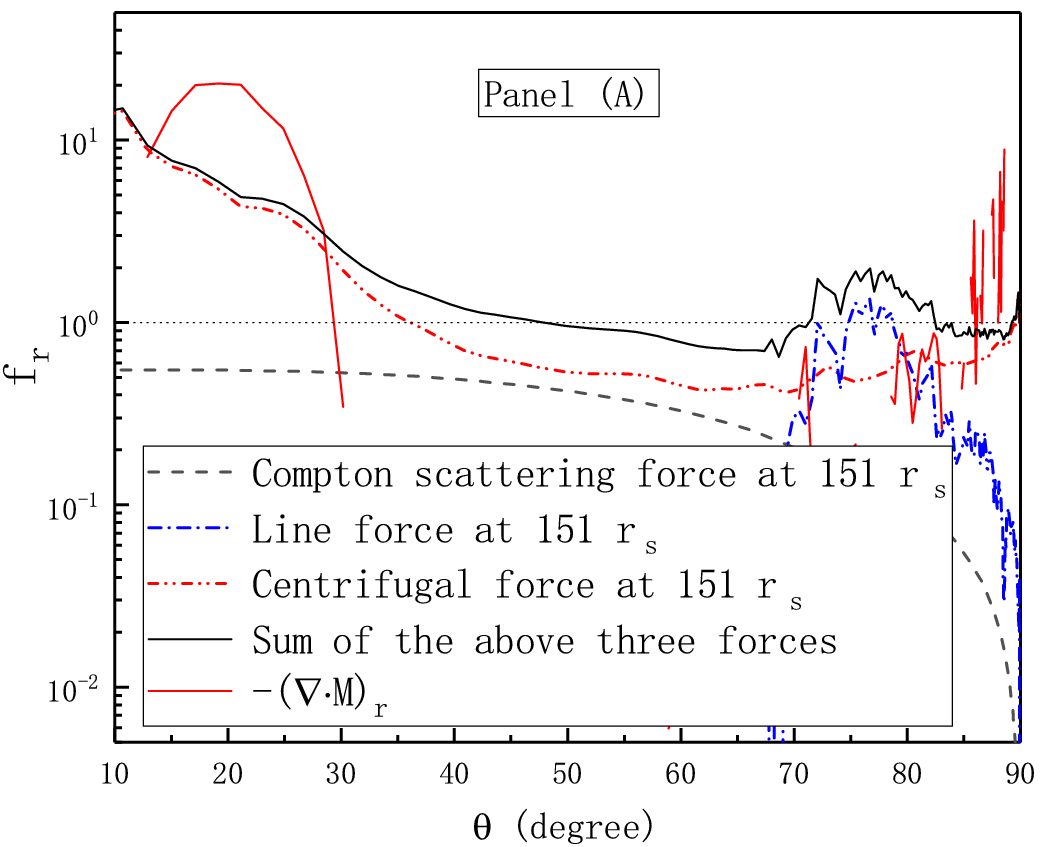}
\includegraphics[width=.45\textwidth]{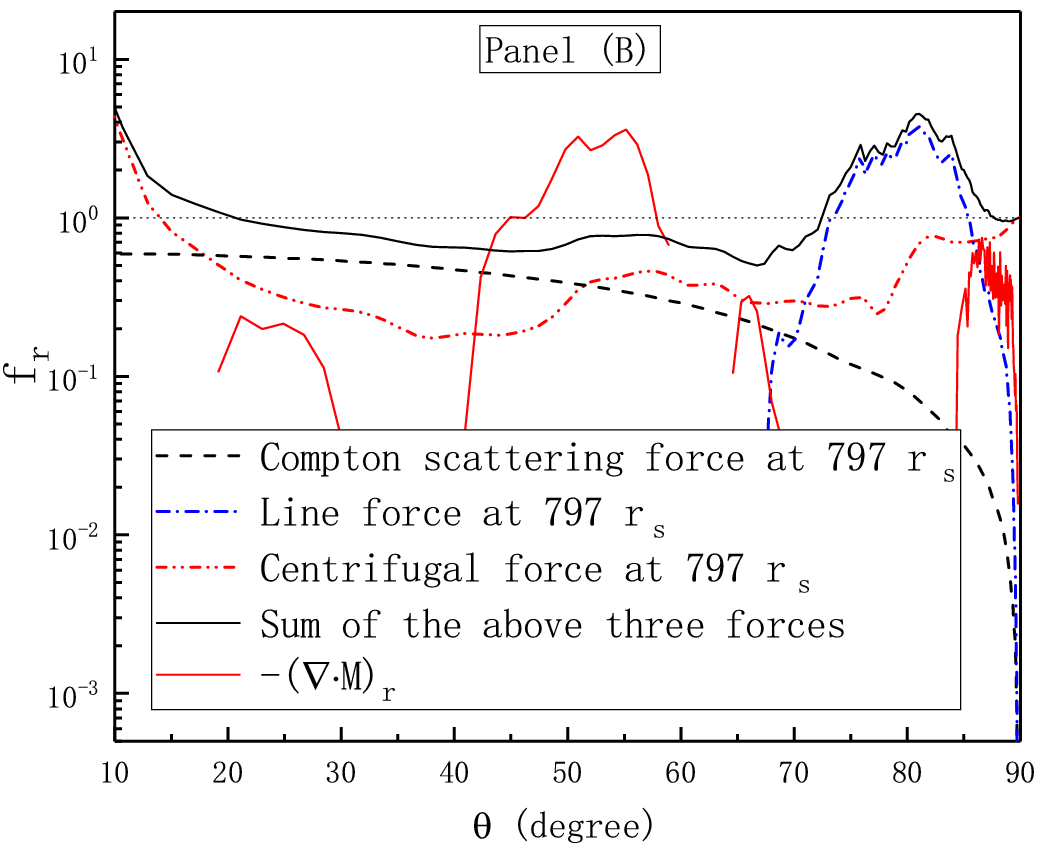}

\ \centering \caption{Angular dependence of radial forces in units of BH gravitation force, for run LDMHD0.6. In figure, dashed lines mean compton-scattering force; dotted-dashed lines means line force; double-dotted-dashed lines mean centrifugal force; solid lines mean $-(\nabla\cdot \bf{M})_{\rm r}$, where $\bf{M}$ is the magnetic stress tensor. Panels A and B show the radial forces at 151 and 797 $r_{\rm s}$, respectively. }
\label{fig8}
\end{figure}

Magnetic fields play an important role in driving high-velocity winds and jets. The radio jets are commonly detected in radio-loud AGNs and radio galaxies. One believes that the formation and acceleration of jets are often attributed to an ordered large-scale magnetic field (e.g. Hada et al. 2011; Nakamura et al. 2018; Chen \& Zhang 2021). Blandford \& Payne (1982) suggested a magneto-centrifugally driven mechanism called the BP model, where an ordered large-scale magnetic field is required and a poloidal component ($B_{\rm p}$) is also required to be stronger than or comparable to the toroidal component ($B_{\phi}$), i.e. $|B_{\phi}/B_{\rm p}|\lesssim1$. When the toroidal magnetic field is much stronger than the poloidal component, i.e. $|B_{\phi}/B_{\rm p}|\gg1$, magnetic pressure plays an important role in driving winds. We show the strength of magnetic fields and the field lines on the $r$--$z$ plane in Figure \ref{fig4}. As shown in panels A and B, magnetic pressure is stronger than the gas pressure in most of the computational region, and the magnetic pressure is almost dominated by the toroidal magnetic field. Especially, in the region of $\theta=\sim10^{\rm o}$--$70^{\rm o}$, where high-velocity winds are formed, the toroidal magnetic field is at least ten times stronger than the poloidal component, and the magnetic pressure is one to tens times stronger than the gas pressure. This is related to the initial strength of magnetic fields in run LDMHD0.6, where the ratio of magnetic pressure to gas pressure at the disk surface is set to be factor of 10 at the inner boundary. As shown in panel C, the field lines become more tilted, compared with the initial field. This is helpful to load materials on field lines from the cold disk surface. In the BP model, the magneto-centrifugally driven outflows are strongly influenced by the inclination angle of the field lines at the cold disk surface. Materials can be loaded on the field lines from the cold disk surface by the magneto-centrifugal force, only when the field lines are tilted ($<60^{o}$ with respect to the disk surface).

\begin{figure}
\includegraphics[width=.45\textwidth]{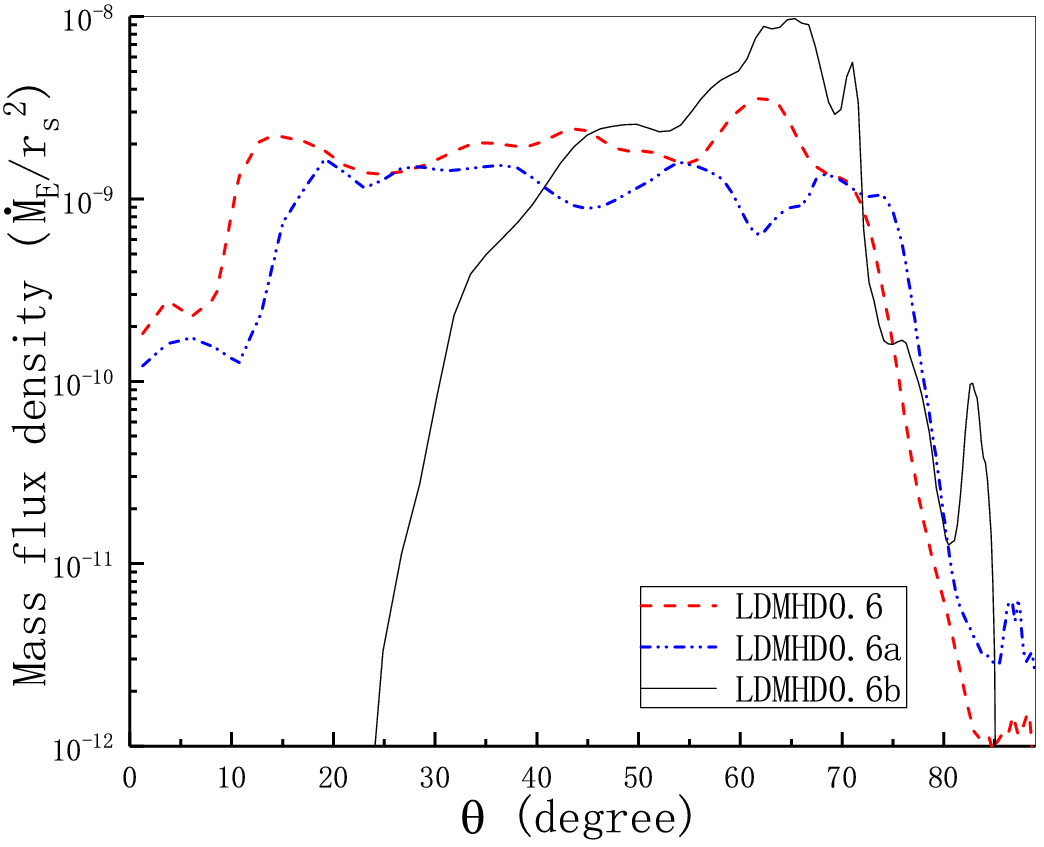}

\ \centering \caption{Angular distribution of the mass flux density at the outer boundary for runs LDMHD0.6, LDMHD0.6a, and LDMHD0.6b.}
\label{fig9}
\end{figure}

\begin{figure}
\includegraphics[width=.45\textwidth]{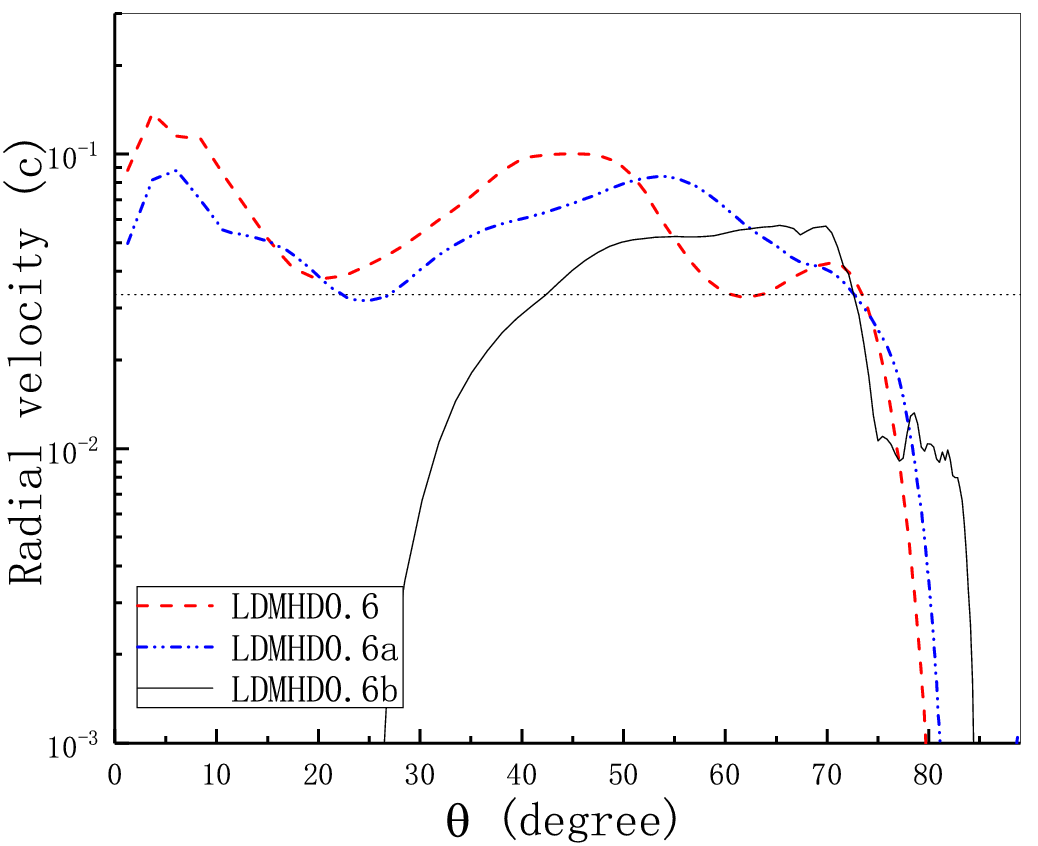}

\ \centering \caption{Angular distribution of the outflow velocity at the outer boundary for runs LDMHD0.6, LDMHD0.6a, and LDMHD0.6b. The dotted line means the velocity of $10^4$ km$\cdot$s$^{-1}$.}
\label{fig10}
\end{figure}

\subsection{Analysis of the forces exerted on winds} \label{subsec:Analysis of Forces}

In run LDMHD0.6, the high-velocity winds are distributed over $\sim10^{\rm o}<\theta<\sim78^{\rm o}$, as shown in Figure \ref{fig4}. Insides 300 $r_{\rm s}$ is the formation region of winds while outsides 300 $r_{\rm s}$ is the propagation region of winds, as shown in panel A of Figure \ref{fig5}.  To understand the formation and acceleration of winds, Figures \ref{fig7} and \ref{fig8} show the angular dependence of radial and angular forces at different radii, respectively. For the formation and acceleration region, we take the forces at 151 $r_{\rm s}$ as an example, as shown in panels A of Figures \ref{fig7} and \ref{fig8}. For the propagation region, we take the forces at 797 $r_{\rm s}$ as an example, as shown in panels B of Figures \ref{fig5} and \ref{fig6}.

In Figure \ref{fig7}, when a quantity is less than zero, the force is toward the rotational axis. Panel A in Figure \ref{fig7} shows that line force blows materials away from the thin disk at small radii and an angular Lorentz force ($-(\nabla\cdot\bf{M})_{\theta}$, where $\bf{M}$ is the magnetic stress tensor) is also helpful to drive the materials away from the thin disk surface. At $\theta<70^{\rm o}$, the angular Lorentz force further drives the gases to move toward the high-latitude region. Panel B also indicates that the angular Lorentz force plays the same role at large radii when $\theta<80^{\rm o}$. The angular Lorentz force is dominated by the gradient force of magnetic pressure.

\begin{figure*}
\includegraphics[width=.32\textwidth]{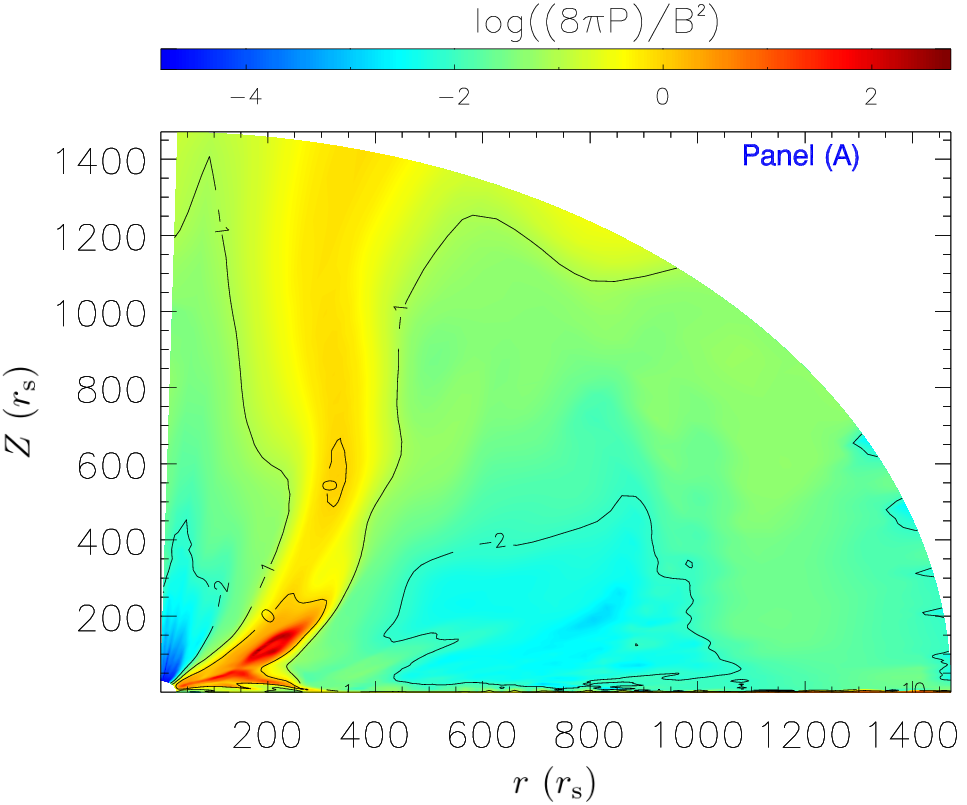}
\includegraphics[width=.32\textwidth]{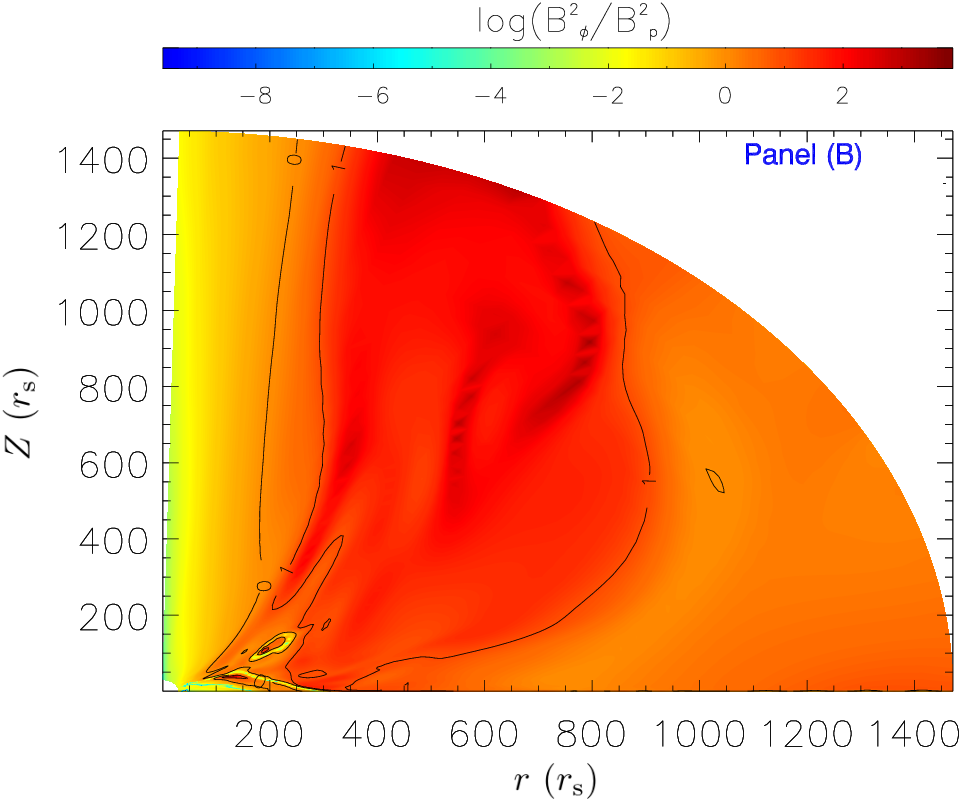}
\includegraphics[width=.32\textwidth]{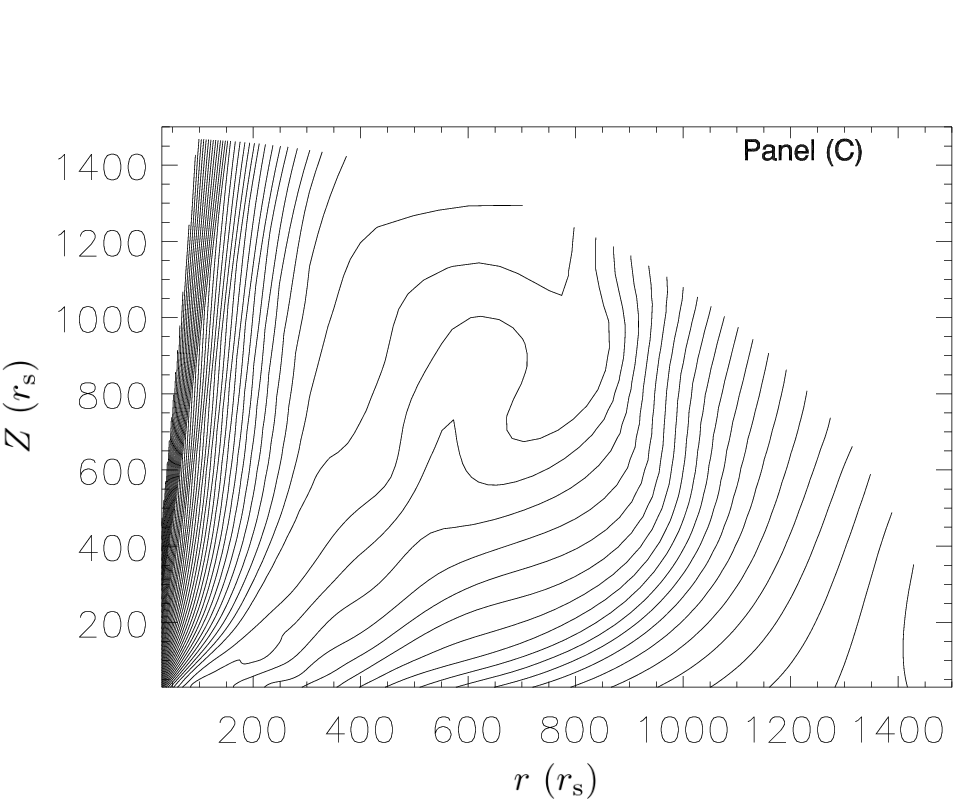}

\ \centering \caption{Magnetic field structure of a pure MHD model. Panels A--C are the same as the panels in Figure \ref{fig4}.}
\label{fig11}
\end{figure*}

\begin{figure*}
,\includegraphics[width=.32\textwidth]{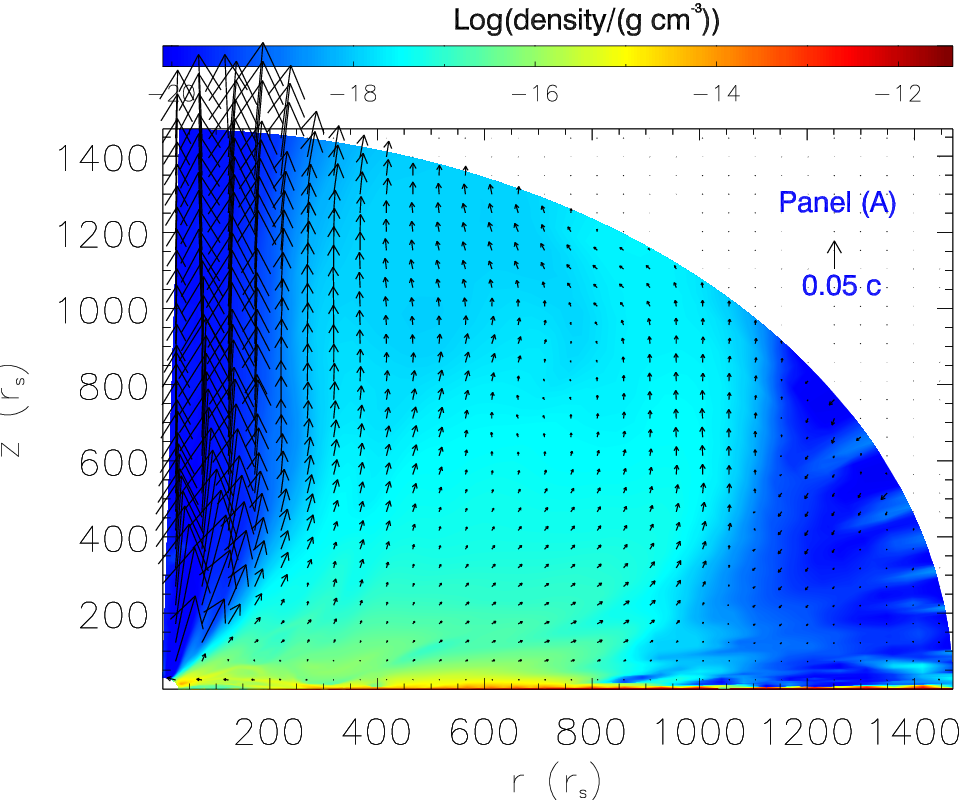}
\includegraphics[width=.32\textwidth]{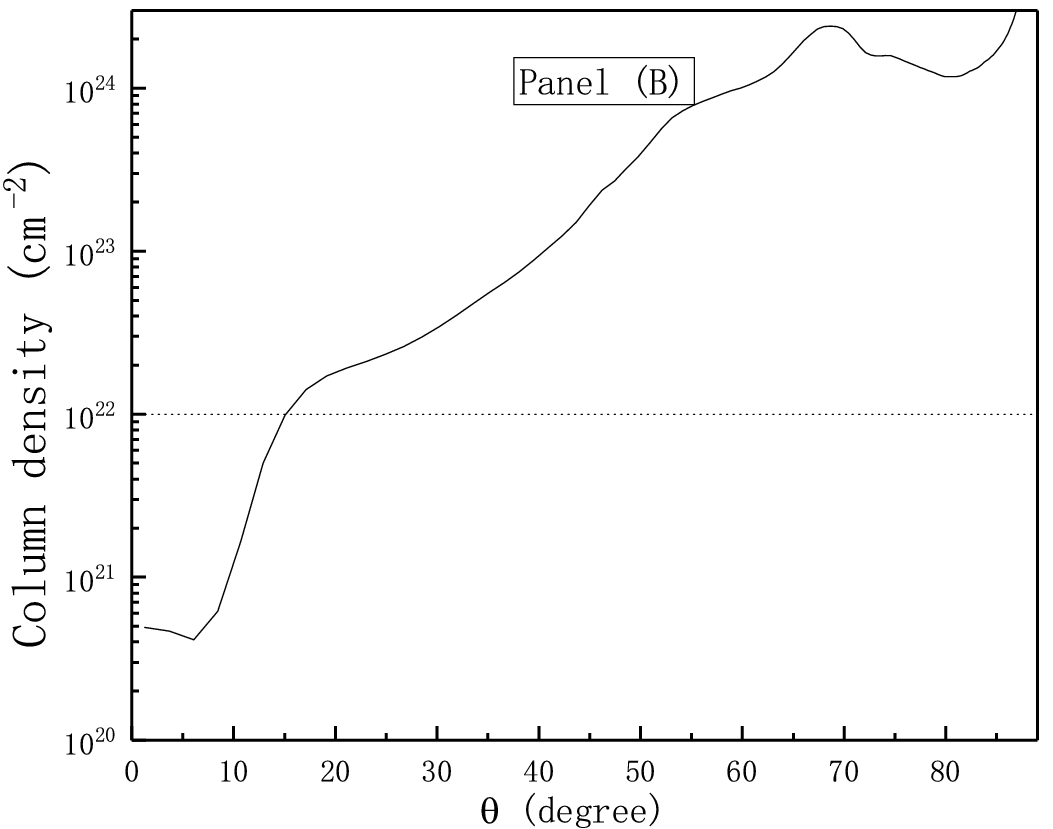}
\includegraphics[width=.32\textwidth]{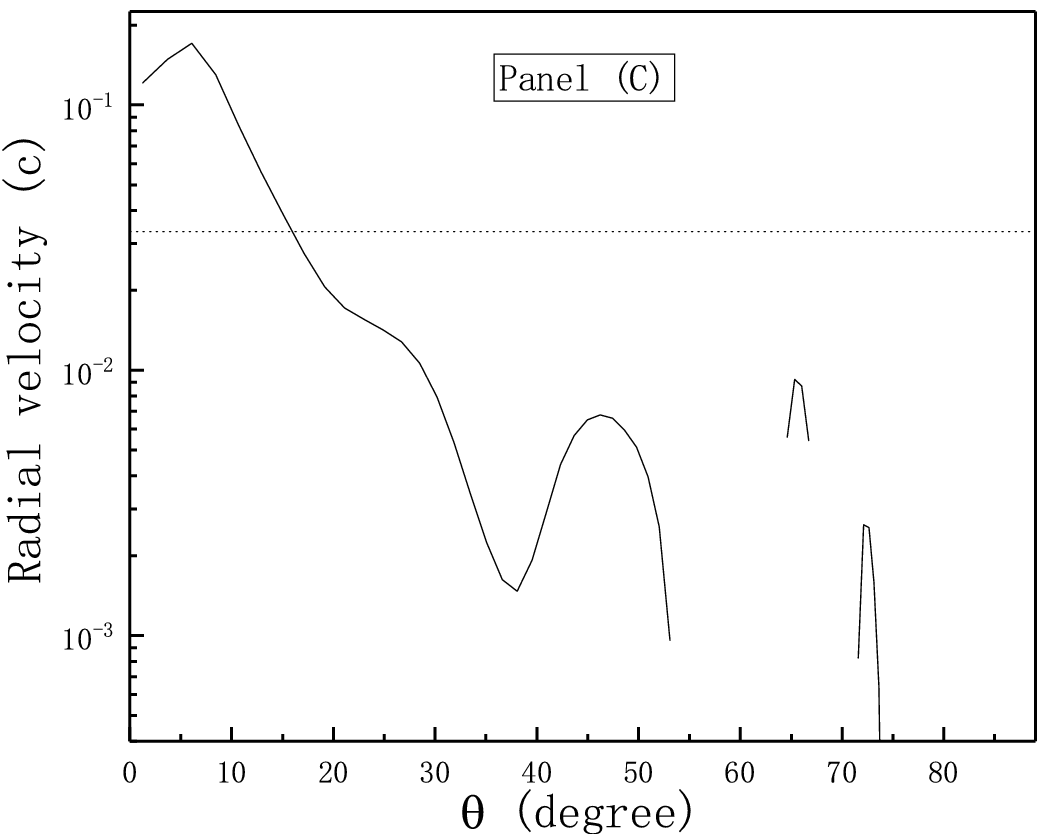}

\ \centering \caption{Basic properties of a pure MHD model. Panel A: density distribution and poloidal velocity; panel B: the angular dependence of column density; panel C: the angular dependence of radial velocity at the outer boundary. The dotted line in panel B means the column density of $10^{22}$ cm$^{-2}$. The dotted line in panel C means the velocity of $10^4$ km$\cdot$s$^{-1}$.}
\label{fig12}
\end{figure*}

In Figure \ref{fig8}, when a quantity is larger than 1, the force is stronger than the BH gravitational force. If the value of a force is larger than zero, the force points outward. In Figure \ref{fig8}, we employ logarithmic coordinates on the \textit{Y}-axis. When the value of a force is less than 0, the curve appears to be intermittent. Panel A shows that the radial component of centrifugal force is stronger than the BH gravitational force over $\theta<\sim35^{\rm o}$. Over $\sim35^{\rm o}<\theta<\sim48^{\rm o}$, the sum of centrifugal force and Compton-scattering force is stronger than the BH gravitational force, though the centrifugal force is weaker than the BH gravitational force. Over $\sim48^{\rm o}<\theta<90^{\rm o}$, the sum of centrifugal force and Compton-scattering force is weaker than the BH gravitational force. Over $\sim70^{\rm o}<\theta<\sim82^{\rm o}$, the sum of centrifugal force and line force is almost stronger than the BH gravitational force. The radial component ($-(\nabla\cdot\bf{M})_{\rm r}$) of Lorentz force also significantly contributes to accelerating outflows at $\sim12^{\rm o}<\theta<\sim30^{\rm o}$. Panel B shows that only when $\theta<\sim14^{\rm o}$ the radial centrifugal force is stronger than the BH gravitational force while when $\theta>\sim14^{\rm o}$ the radial centrifugal force is weaker than the BH gravitational force. Over $\sim73^{\rm o}<\theta<\sim85^{\rm o}$, the radial component of line force is almost stronger than the BH gravitational force. The strength of radial line force increases with the increase of radius at some angles. At small radii, such as 151 $r_{\rm s}$, the angular line force is a dominant component of line force over $80^{\rm o}<\theta<90^{\rm o}$ to blown away materials from the disk surface. At large radii, such as 797 $r_{\rm s}$, the radial line force is an important force to drive winds. Over $\sim45^{\rm o}<\theta<\sim58^{\rm o}$, the radial component ($-(\nabla\cdot\bf{M})_{\rm r}$) of Lorentz force is stronger than gravitational force and then significantly contributes to accelerating winds.

If magnetic fields are not included in simulations, gases keep angular momentum constant along streamlines and then the centrifugal force decays faster than the BH gravitational force. When magnetic fields are included in simulations, the effect of magnetic fields could contribute to the enhancement of centrifugal force. At $\theta<\sim30^{\rm o}$, the enhancement of centrifugal force is attributed to the effect of magnetic fields. This force is a magneto-centrifugal force. To confirm whether the gas angular momenta at $\theta>35^{\rm o}$ increase due to the effect of magnetic fields, we track streamlines on the $r$--$z$ plane (as shown in Figure \ref{fig5}) and compute the angular momenta at 151 $r_{\rm s}$ and a starting position along the streamlines. The starting position of the streamlines is close to the disk surface. We find that, when gases flow from the starting position to 151 $r_{\rm s}$ along streamlines, the gas angular momenta could increase while the biggest increase is not more than 15\% over $35^{\rm o}<\theta<60^{\rm o}$. This insures that the biggest increase of centrifugal force is not more than 2.5\%. Therefore, the magneto-centrifugal force, caused by magnetic fields, is negligible at $\theta>35^{\rm o}$.

Here, we summarize the formation and acceleration of winds as follows. The line force pushes up materials away from the disk surface inside 300 $r_{\rm s}$ and accelerates them over $70^{\rm o}<\theta<90^{\rm o}$. The angular Lorentz force can drive the gases at $\theta<70^{\rm o}$ to further move toward the high-latitude region. This is the key to form the low-inclination winds. At $\theta<30^{\rm o}$, the magneto-centrifugal force is an important force of driving winds while it becomes weaker at large radii. At $30^{\rm o}<\theta<60^{\rm o}$, both the centrifugal force and the Compton-scattering force balance the BH gravitational force at some degree while the two forces at large radii become weak in balancing the BH gravitational force. At $\sim73^{\rm o}<\theta<\sim85^{\rm o}$, line force is always effective in driving the winds, especially at large radii. At large radii, the radial Lorentz force plays an important role in accelerating winds at some angles.

\subsection{Effects of magnetic field strength} \label{subsec:Effect of magnetic field strength}

We use runs LDMHD0.6, LDMHD0.6a, and LDMHD0.6b in Table 1 to test the effects of magnetic strength. From runs LDMHD0.6 to LDMHD0.6a, where $\beta_{0}$ increases from 0.1 to 0.2, the mass outflow rates decrease due to the initial magnetic field becoming weak. For run LDMHD0.6b, $\beta_{0}=100.0$, i.e. the magnetic field is very weak. Compared with runs LDMHD0.6 and LDMHD0.6a, however, the mass outflow rate of run LDMHD0.6c is higher than that of runs LDMHD0.6 and LDMHD0.6a. As shown in Figure \ref{fig9}, for runs LDMHD0.6 and LDMHD0.6a, the angular distributions of their mass flux density are similar to each other. Compared with run LDMHD0.6, the mass flux density of run LDMHD0.6a is slightly low. However, the mass flux density of run LDMHD0.6b is much lower than that of runs LDMHD0.6 and LDMHD0.6a when $\theta<40^{\rm o}$. The reasons are as follows. As discussed in section 3.3, the diffusing of gases from low-latitude regions to high-latitude regions is driven by the angular component ($-(\nabla\cdot\bf{M})_{\theta}$) of Lorentz force. A weak magnetic field, such as the case of run LDMHD0.6b, weakens the diffusing of gases toward high-latitude regions. For run LDMHD0.6b, its magnetic field is much weaker than that of runs LDMHD0.6 and LDMHD0.6a, which means that $-(\nabla\cdot\bf{M})_{\theta}$ can be neglected in run LDMHD0.6b and then gases cannot further diffuse toward the high-latitude region of $\theta<40^{\rm o}$. Therefore, high-velocity winds cannot form at $\theta<40^{\rm o}$ in run LDMHD0.6b. From Figure \ref{fig9}, we can see that the mass flux density of run LDMHD0.6b is much lower than that of the run models at $\theta<40^{\rm o}$. However, at $\theta=50^{\rm o}$--70$^{\rm o}$, the mass outflow rate of run LDMHD0.6b is higher than that of the two runs, which causes the mass outflow rate of run LDMHD0.6b to be higher than that of the two runs (as given in Table 1). This is because winds cannot effectively diffuse toward the high-latitude region in the case of weak magnetic fields and then the shielding of X-ray photons becomes effective. Line-driving winds is more effective in the case of weak magnetic fields (such as run LDMHD0.6b) than in the case of strong magnetic fields (such as run LDMHD0.6). This is helpful to blow more materials away from the disk surface in the case of weak magnetic fields. In the case of strong magnetic fields, the effect of X-ray shielding for line-driving winds is reduced due to the diffusion of materials toward the higher-latitude region ($\theta<40^{\rm o}$).

Figure \ref{fig10} shows the angular profiles of the outflow velocity at the outer boundary. For runs LDMHD0.6 and LDMHD0.6a, their velocity profiles are basically similar. In the case of weak magnetic field, such as run LDMHD0.6b, the high-velocity winds can not be observed at the high-latitude region (i.e. $\theta<\sim40^{\rm o}$). Comparing run LDMHD0.6 and run LDMHD0.6b, their velocity angular profiles at the outer boundary is significantly different. At $\theta=70^{\rm o}$, the radial velocity at the outer boundary in run LDMHD0.6 is reduced due to reducing the effect of X-shielding by the diffusion of gases toward higher latitudes. At $\theta=50^{\rm o}$, the radial velocity at the outer boundary in run LDMHD0.6 is increased due to the accelerating by the radial component of Lorentz force (see panel B in Figure 8). In the case of weak magnetic field, outflows have higher velocity at the low-latitude region ($\sim70^{\rm o}$) than in case of strong magnetic fields. At the middle- and high-latitude region ($<60^{\rm o}$), outflows have higher velocity in the case of strong magnetic fields than in the case of weak magnetic fields. Except for the effects of strong magnetic fields, there does not seem to be another mechanism to form the high-velocity winds at the high-latitude region ($<40^{\rm o}$).

In summary, comparing the two runs of $\beta_0=0.1$ and 0.2, the strength of magnetic fields cannot significantly change the angular distribution of winds, but slightly affects the mass outflow rate. Runs LDMHD0.6 and LDMHD0.6a are classified in the case of strong magnetic field (i.e. low $\beta$), while runs LDMHD0.6b is classified in the case of weak magnetic field (i.e. high $\beta$). Comparing the two cases, the angular distribution of winds is different while the other properties (such as the wind velocity, the ionization parameter, and the column density) are basically similar.

The UFOs in radio-loud AGNs are measured at $\theta=\sim10^{\rm o}$--$\sim70^{\rm o}$ (Tombesi et al. 2014), which indicates that the UFOs in radio-loud AGNs have a broad opening angle. The UFOs are detected in more than 35\% of radio-quiet AGNs (Tombesi et al. 2011). If the detection probability of UFOs in radio-quiet AGNs is 35\%, the UFOs in radio-quiet AGNs could have a narrow opening angle. In the present simulations, the high-velocity winds in the low-$\beta$ case have a broader opening angle than that in the high-$\beta$ case. From this point of the formation of UFOs, it is conjectured that the UFOs in radio-loud AGNs may be attributed to the low-$\beta$ case while the UFOs in radio-quiet AGNs may be attributed to the high-$\beta$ case. In the radio-loud AGNs, the outflow morphology could be different from that in the radio-quiet AGNs. However, the observation of the outflow morphology is difficult. The UFO morphology in radio-loud AGNs and radio-quiet AGNs has not been directly measured. Therefore, there is no more direct evidence to support the conjecture.

In addition, run LDMHD0.6b in Table 1 is the same as run RMHDW0.6 in Yang et al. (2021), but run LDMHD0.6b has slightly lower mass outflow rates and outflow velocities than run RMHDW0.6. This is caused by employing different time intervals when time-averaging is implemented. In Yang et al. (2021), the time interval of averaging is from 0.5 $T_{\rm orb}$ to 1.0 $T_{\rm orb}$, while here the time interval is from 0.089 $T_{\rm orb}$ to 0.127$T_{\rm orb}$. Figure 2 in Yang et al. (2021) shows that the mass outflow rate over 0.089--0.127 $T_{\rm orb}$ is less than that over 0.5--1.0 $T_{\rm orb}$ for run RMHDW0.6.

\subsection{Comparison to pure MHD simulation} \label{subsec:Comparison with pure MHD simulation}

For comparison with previous simulations, we implement a pure MHD simulation. Except for that radiation force is not included in the pure MHD simulation, other parameters are the same as those in run LDMHD0.6c. After a quasi-steady state is reached, we time average the 250 snapshots of simulation over the time interval of 0.234--0.272 $T_{\rm orb}$. Figures \ref{fig11} and \ref{fig12} show the time-averaged solutions. Figure \ref{fig11} shows magnetic field structure. As shown in Figure \ref{fig11}, magnetic pressure is dominant in most of the computational domain and the toroidal magnetic field is stronger than the poloidal field in the dense winds. The poloidal magnetic field becomes more tilted than the initial field. Panel A of Figure \ref{fig12} shows density distribution and poloidal velocity. The basic structure of outflows is very different from that in panel A of Figure \ref{fig5}. In this paper, we aim at explaining UFOs in radio-loud AGNs. Therefore, we focus on the winds that have column density of higher than $10^{22}$ cm$^{-2}$ and velocity of higher than 10$^4$ km s$^{-1}$. Panel B shows that when $\theta>\sim15^{\rm o}$ the column density is higher than $10^{22}$ cm$^{-2}$, while the winds over $\theta>\sim15^{\rm o}$ have velocity of less than 10$^4$ km s$^{-1}$. This seems to imply that the pure MHD simulation cannot be used to explain the formation of UFOs. When magnetic fields are stronger, it is necessary to check whether UFOs could be formed or not. However, the strength of magnetic fields can not be further strengthened in our simulating, because our simulations crashes due to that time step becomes much shorter than that in current simulations. The case of stronger magnetic fields is worth to be studied again in the future based on numerical simulations.

\section{Discussion} \label{sec:Dissussion}
In the present simulations, the effects of X-ray spectral shape are not taken into account. However, the X-ray spectral shape may have an impact on line-driving winds via influencing the ionization degree of materials. In observations, it is suggested that the column density and velocity of UFOs are correlated with not only the X-ray luminosity but also the X-ray hardness (Chartas et al. 2009; Matzeu et al. 2017). For example, in PDS 456, the UFO velocity is observed to be a positive correlation with the X-ray strength. This correlation seems to be interpreted as an evidence for both a radiatively driven wind (Matzeu et al. 2017) and a magnetically driven winds (Fukumura et al. 2018). It is not clear whether the increase of X-ray luminosity in PDS 465 is accompanied by the increase of UV luminosity. According to the line-driving wind model, when the UV luminosity keeps constant, the X-ray luminosity increases, the ionization parameter of winds could increase, and then the wind velocity decreases due to the weakening of line force. However, if the UV luminosity increases with the X-ray luminosity in PDS 456, the wind velocity could increase. On the other hand, Fukumura et al. (2018) applied a self-similar MHD-driven wind model to reproduce the observed correlation of the UFO velocity with the X-ray strength of PDS 456. The launching mechanism of the UFOs in PDS 465 is still a debatable issue.

The hardness of X-ray photons from the corona has not been taken into account in the present simulations. The present simulations have ignored the interaction between the corona X-ray photons and the wind materials. Due to scattering and absorption, the X-ray spectra evolute with time. Local X-ray spectra determine the local ionization parameter ($\xi$) which is very important to determine the strength of line force and define the UFOs in this work as a proxy. However, when the evolution of X-ray spectra is taken into account in the simulations, the HD/MHD simulations of coupling radiative transfer may need to be implemented. This is computationally expensive. Decoupling radiative transfer and HD/MHD calculations becomes an alternative method, called a post-process approach. The post-process approach is often implemented in previous works (e.g. Fukumura et al. 2010; Sim et al. 2010; Higginbottom et al. 2014; Chakravorty et al. 2016). Based on the post-process approach, it is found that the ``shielding'' effect is not necessarily as effective as estimated earlier in the line-force-driven winds, because the scattered X-ray photons are effective in ionizing the materials that are shielded from the X-ray photons from the central source (Sim et al. 2010; Higginbottom et al. 2014). However, real flows above the accretion disk could adjust themselves to make the scattered photons not to be strong enough, so that the ``shielding'' effect could be still effective and then the line force could efficient in driving winds. This is a speculation and the verification of this speculation is beyond the scope of this work.

\section{Summary} \label{sec:Summary}
Tombesi et al. (2014) found that the UFOs in radio-loud AGNs can be detected at the inclination angle of $\sim10^{\rm o}$--$70^{\rm o}$ away from jets. Except that the UFOs in radio-loud AGNs have such low inclination, their basic properties are similar to that in radio-quiet AGNs. The properties of UFOs seem to be explained in both a line-force-driving model and a generic magnetic-driving mechanism. However, the line-force-driving model cannot describe the low-inclination UFOs in radio-loud AGNs, because the line-force-driven winds form at middle and low latitudes.

In this paper, we have simulated the disk winds driven by both radiation forces (including Compton-scattering force and line force) and magnetic field and also implemented the numerical simulation of only magnetically driven winds. We set the strength of initial magnetic fields is to ten times stronger than the gas pressures at the inner boundary of disk surface. As addressed in section 3.1, we define the high-velocity winds as the outflows that have a velocity higher than 10$^4$ km s$^{-1}$ and the column density higher than $10^{22}$ cm$^{-2}$. UFOs are observed in the high-velocity winds, when the hybrid models of magnetic fields and radiation forces are employed. Under the same initial conditions, the hybrid models are more helpful to form UFOs than the pure MHD models. In the hybrid models, the UFOs can be generated at a low-inclination angle. For example, the UFOs in run LDMHD0.6c are distributed in the angular range of $\sim10^{\rm o}$--$\sim75^{\rm o}$, where the moderately ionized gases of log($\xi$/(erg s$^{-1}$ cm))$\sim$3--6 are clumpy and their location changes with time. When the magnetic fields are stronger, we may observe the UFOs with a lower inclination angle (i.e. close to $\sim10^{\rm o}$) from simulations. It could be a choice that the hybrid models are suggested to explain the formation of UFOs.

Based on the hybrid models, Everett (2005) investigated the effect of radiation field on self-similar MHD winds and found the efficiency of a radiatively driven wind increases due to shielding by a magneto-centrifugal wind. In Everett's work (2005), where the disk luminosity is less than or equals 0.1$L_{\rm Edd}$, the role of the radiation field is not important in driving winds, compared with magnetic fields (see Figure 10 and 14 in Everett (2005)). In the present simulations, the magneto-centrifugal force is not important in accelerating winds, but the magnetic fields can drive gases toward the high-latitude region, which weakens shielding and then decreases the efficiency of line-driving winds.

Comparing the low-$\beta$ (such as runs LDMHD0.6 and LDMHD0.6a) and high-$\beta$ (such as run LDMHD0.6b) cases, the wind properties (such as the wind velocity, the ionization parameter, and the column density) are basically similar, except for the angular distribution of winds. The high-velocity winds in the low-$\beta$ case have a broader opening angle than that in the high-$\beta$ case. From this point of the formation of UFOs, we conjecture that the UFOs observed in radio-loud and radio-quiet AGNs may be attributed to the low-$\beta$ case and the high-$\beta$ case, respectively.

\acknowledgments{This work is supported by the Natural Science Foundation of China (grant 11973018) and Chongqing Natural Science Foundation (grant cstc2019jcyj-msxmX0581). The authors thank the anonymous referee for the constructive suggestions.}

\end{document}